\documentclass[12pt]{article}
\usepackage{amssymb}
\usepackage{amsfonts}
\usepackage{amsmath}
\usepackage[mathscr]{eucal}
\usepackage{amssymb}
\usepackage{amsthm}
\usepackage{bbold}
\usepackage{bm}
\usepackage{graphicx}
\usepackage{caption}

\theoremstyle{plain}

\numberwithin{equation}{section}
\newcommand{\be}{\begin{equation}}
\newcommand{\ee}{\end{equation}}
\newcommand{\bea}{\begin{eqnarray}}
\newcommand{\eea}{\end{eqnarray}}
\newcommand{\nn}{\nonumber \\}
\newcommand{\lb}{\label}
\newcommand{\p}[1]{(\ref{#1})}

\begin{document}

\begin{titlepage}

\vspace*{0.2cm}

\renewcommand{\thefootnote}{\star}
\begin{center}

{\LARGE\bf  Quantum SU(2$|$1) supersymmetric}\\

\vspace{0.5cm}

{\LARGE\bf Calogero--Moser spinning systems }\\

\vspace{1.5cm}
\renewcommand{\thefootnote}{$\star$}

{\large\bf Sergey\,Fedoruk}${}^{a}$,\,\,\,{\large\bf Evgeny\,Ivanov}${}^a$,\,\,\,{\large\bf Olaf\,\,Lechtenfeld}${}^b$,\,\,\,{\large\bf Stepan\,Sidorov}${}^a$
 \vspace{0.8cm}

${}^a${\it Bogoliubov Laboratory of Theoretical Physics, JINR, 141980  Dubna, Moscow region, Russia} \\
\vspace{0.2cm}

{\tt fedoruk@theor.jinr.ru, \, eivanov@theor.jinr.ru, \, sidorovstepan88@gmail.com }\\

\vspace{0.6cm}

${}^b${\it Institut f{\" u}r Theoretische Physik and Riemann Center for Geometry and Physics \\
Leibniz Universit{\" a}t Hannover, Appelstrasse 2, 30167 Hannover, Germany} \\
\vspace{0.2cm}
{\tt lechtenf@itp.uni-hannover.de}
\vspace{0.7cm}

\end{center}
\vspace{0.2cm} \vskip 0.6truecm \nopagebreak

\begin{abstract}
\noindent
${\rm SU}(2|1)$ supersymmetric multi-particle quantum mechanics
with additional semi-dynamical spin degrees of freedom is considered.
In particular, we provide an $\mathcal{N}{=}\,4$ supersymmetrization of
the quantum ${\rm U}(2)$ spin Calogero--Moser model, with an
intrinsic mass parameter coming from the centrally-extended
superalgebra $\widehat{su}(2|1)$. The full system admits an
${\rm SU}(2|1)$ covariant separation into the center-of-mass sector
and the quotient. We derive explicit expressions for the classical and
quantum ${\rm SU}(2|1)$ generators in both sectors as well as for
the total system, and we determine the relevant energy spectra,
degeneracies, and the sets of physical states.

\end{abstract}

\vspace{1cm}

\phantom{Keywords: }

\newpage

\end{titlepage}

\setcounter{footnote}{0}

\setcounter{equation}0
\section{Introduction}

The many-particle Calogero-Moser systems \cite{Calogero69a,Calogero69b,Calogero71,Moser,OP} and their generalizations
occupy a distinguished  place in the contemporary theoretical and mathematical physics.
Apart from such notable mathematical properties, as the classical and quantum integrability,
these systems possess a wide range of physical applications which are hard to enumerate.
Among these  applications, it is worth to mention, e.g., a close connection between
the algebra of observables in the Calogero system and the higher-spin algebra,
as was pointed out in \cite{BHV,BHKV}.
In the same papers, there was revealed an important role of Calogero-like models for describing
particles with fractional statistics.
Another widely known applications of the Calogero-Moser systems concern the black hole physics.
It was suggested \cite{GibTow} that the Calogero-Moser systems can provide a microscopic description
of the extreme Reissner-Nordstr\"{o}m black hole in the near-horizon limit.
It was argued that, from the M-theory perspective, an important role in this correspondence should be played
by various ${\cal N}{=}\,4$ supersymmetric extensions of the Calogero-Moser models.
Supersymmetric  Calogero-Moser systems  have also further applications in string theory (see, for example, \cite{MMVer})
and ${\cal N}{=}\,4$ super Yang-Mills theory \cite{Dabh,AP}.

Keeping in mind these physical and mathematical motivations, it seems of great interest to construct and study new
versions of supersymmetric Calogero-type systems.

In a recent paper \cite{FI16}, there was proposed the superfield matrix model of ${\rm SU}(2|1)$ supersymmetric
mechanics\footnote{This kind of deformed ${\cal N}{=}\,4$ supersymmetric mechanics was introduced and studied in
\cite{Sm,BellNer,Romelsberger,IS14a,IS14b,IS15}. Some of the ${\rm SU}(2|1)$ models can be derived by a dimensional reduction from ${\cal N}{=}\,1$ Lagrangians on the curved $d=4$ manifold
$\mathbb{R} \times S^3$ \cite{Casimir,Denef}.}
as a new ${\cal N}{=}\,4$ extension of $d=1$ Calogero--Moser
multi-particle system.
This matrix model is a massive generalization of the multiparticle ${\cal N}{=}\,4$ model constructed and studied in \cite{FIL08,FIL12}. It is naturally formulated
in $d=1$ harmonic superspace \cite{IvLech} and is described by the following set of ${\cal N}{=}\,4$ harmonic superfields:
\begin{itemize}
\item
$n^2$ commuting general superfields $\mathscr{X}_b^a=(\widetilde{\mathscr{X}_a^b})$, $a,b=1,\ldots ,n$
combined into  an hermitian $n{\times}n$-matrix superfield
$ \mathscr{X}=(\mathscr{X}_a^b)$ which transform in adjoint representation of ${\rm U}(n)$ and represent off-shell ${\rm SU}(2|1)$ multiplets $({\bf 1, 4, 3})$;
\item
$n$ commuting analytic complex superfields ${\mathcal{Z}}{}^{+}_a$ forming ${\rm U}(n)$ spinor $\mathcal{Z}^+=(\mathcal{Z}^+_a)$,
$\widetilde{\mathcal{Z}}^{+}=(\widetilde{\mathcal{Z}}^{+a})$ and representing off-shell ${\rm SU}(2|1)$ multiplets $({\bf 4, 4, 0})$;
\item
$n^2$ non-propagating analytic ``topological'' gauge superfields $ V^{++}=(V^{++}{}_a^b)$, $(\widetilde{V^{++}{}_a^b}) =V^{++}{}_b^a$.
\end{itemize}

The matrix superfield $\mathscr{X}=\mathscr{X}\left(\zeta_H\right)$ is defined on the ${\rm SU}(2|1)$ harmonic superspace
$\zeta_H\equiv\left(t_{A}, \theta^\pm, \bar\theta^\pm, w^\pm_i\right)$ ($i=1,2$), while the analytic superfields ${\mathcal{Z}}{}^{+}$,
$\widetilde{\mathcal{Z}}^{+}$ and $V^{++}{}$ on the analytic harmonic subspace $\zeta_{A} =\left(t_{A}, \bar{\theta}^{+}, \theta^{+},  w^{\pm}_i\right) \subset \zeta_H\,$.
The relevant superfield action is written as
\begin{equation}\label{4N-gau}
S_{matrix} =-\frac{1}{4}\int \mu_H  {\rm Tr}  \left(\mathscr{X}^{\,2}\right)
+\frac{1}{2}\int \mu^{(-2)}_A  \mathcal{V}_0
\widetilde{\mathcal{Z}}{}^{+a} \mathcal{Z}^+_a
+\frac{i}{2}\,c\int \mu^{(-2)}_A  \,{\rm Tr} \,V^{++}\,,
\end{equation}
where the invariant integration measures are written as
\be
    \mu_H = dw\, dt_{A} \,d\bar{\theta}^{-}d\theta^- d\bar{\theta}^{+}d\theta^{+}
\left(1 +m\, \theta^{+}\bar{\theta}^{-} - m\,\theta^{-}\bar{\theta}^{+} \right),\qquad
    \mu^{(-2)}_A = dw\, dt_{A} \,d\bar{\theta}^{+}d\theta^{+}\,.
\ee
The mass-dimension  parameter $m$ is encoded in the centrally-extended superalgebra $\widehat{su}(2|1)$ as the contraction parameter to the flat
${\cal N}{=}\,4, d=1$ superalgebra.\footnote{It was shown in \cite{IS14b}, that the centrally extended  superalgebra $\widehat{su}(2|1)$
can be represented
as a semi-direct sum of $su(2|1)$ and an extra R-symmetry generator: $\widehat{su}(2|1)\simeq su(2|1) {+\!\!\!\!\!\!\supset}\, u(1)$. The central charge is a combination
of the R-symmetry generator and the internal ${\rm U}(1)$ generator of $su(2|1)$. In the models under consideration the central charge operator is identified
with the canonical Hamiltonian.}
It does not explicitly appear in \p{4N-gau} but comes out in the component action from the measure $\mu_H$ and the $\theta$-expansion of the superfields $\mathscr{X}$
as a result of solving the appropriate ${\rm SU}(2|1)$ covariant constraints (for more details, see \cite{FI16, IS14a}).The local ${\rm U}(n)$ transformations of the involved superfields are given by
\begin{equation}\label{tran4}
\mathscr{X}^{\,\prime} =  e^{i\lambda} \mathscr{X} e^{-i\lambda} , \qquad
\mathcal{Z}^+{}^{\prime}
= e^{i\lambda} \mathcal{Z}^+ , \qquad
V^{++}{}^{\,\prime} =  e^{i\lambda}\, V^{++}\, e^{-i\lambda} - i\, e^{i\lambda} (D^{++}
e^{-i\lambda})\,.
\end{equation}
The superfield $\mathcal{V}_0\left(\zeta_A\right)$ is a prepotential for the singlet part ${\rm Tr} \left( \mathscr{X} \right)$
of the matrix superfield $\mathscr{X}$ (see details in \cite{FI16}).
The constant $c$ in (\ref{4N-gau}) is a parameter of the model. After quantization, it specifies external ${\rm SU}(2)$ spins of the physical
states, $c \,\rightarrow \, 2s + 1\in \mathbb{Z}_{>0}\,$,  which implies that the set of these states  splits into irreducible ${\rm SU}(2)$ multiplets.

The matrix $d=1$ superfield $\mathscr{X}$ has the physical component fields $X=(X_a{}^b)=X^\dagger$, $\Psi^k=(\Psi^k{}_a{}^b)$ and auxiliary bosonic component fields,
the superfields $ \mathcal{Z}^+_a, \widetilde{\mathcal{Z}}{}^{+a}\,$ have the bosonic components $Z^\prime{}^k=(Z^\prime{}^k_a)$,
$\bar{Z}^\prime{}_k=(\bar{Z}^\prime{}_k^a)=(Z^\prime{}^k)^\dagger$ and auxiliary fermionic fields.  Choosing the WZ gauge
$V^{++} =2i\,\theta^{+} \bar\theta^{+}A(t_A)$,
eliminating auxiliary fields and redefining the spinor fields as $ Z^\prime{}^i_a \rightarrow
Z^i_a/\left({\rm Tr} (X)\right)^{1/2}$, we obtain from (\ref{4N-gau}) the on-shell component action
\begin{eqnarray}\label{4N-gau-bose-a}
S_{matrix}  &=& S_b + S_f,
\\
S_b &=& \frac12\, {\rm Tr}\int dt \,\Big( \nabla X\nabla X - m^2 X^2 \Big)-c\int dt \,{\rm Tr}A
\nonumber \\
&& + \frac{i}{2}\, \int dt \,
\Big(\nabla \bar Z_k \, Z^k-\bar Z_k \nabla Z^k \Big) + \int dt \,
 \frac{S^{(ik)}S_{(ik)}}{4(X_0)^2}\,,\label{4N-gau-bose-1}
\\
  S_f&=&   \frac12\,{\rm Tr} \int dt \Big[i\left( \bar\Psi_k \nabla\Psi^k
-\nabla\bar\Psi_k \Psi^k
\right) +2m \bar\Psi_k \Psi^k\Big]  -\int dt  \,\frac{\Psi^{(i}_0\bar\Psi^{k)}_0 S_{(ik)}}{(X_0)^2}\,.\label{4N-gau-fermi-1}
\end{eqnarray}
Here,
$$
X_0 := \frac{1}{\sqrt{n}}\,{\rm Tr} (X), \qquad\Psi_0^i := \frac{1}{\sqrt{n}}\, {\rm Tr} (\Psi^i),
\qquad\bar\Psi_0^i := \frac{1}{\sqrt{n}}\,
{\rm Tr} (\bar\Psi^i)\,,
$$
\begin{equation}\label{def-S}
S_{(ik)}:= \bar Z_{(i}Z_{k)} := \bar Z_{(i}^a Z_{k)a}\,,
\end{equation}
and $(\nabla \bar Z_k \, Z^k):= \nabla \bar Z_k^a \, Z^k_a$.
The ${\rm U}(n)$ gauge-covariant
derivatives in (\ref{4N-gau-bose-1}), (\ref{4N-gau-fermi-1}) are defined by
\begin{equation}\label{cov-X-M}
\nabla X =
\dot X +i[ A,X] \,,
\qquad \nabla \Psi^{i} =
\dot\Psi^{i} +i[ A,\Psi^{i}] \,,\qquad  \nabla \bar\Psi_{i} =
\dot{\bar\Psi}_{i} +i[ A,\bar\Psi_{i}]\,.
\end{equation}
\begin{equation}\label{cov-Z-M}
\nabla Z^k=\dot Z^k + i A \, Z^k\,, \qquad\nabla \bar Z_k=\dot{\bar Z}_k -i\bar Z_k\, A\,.
\end{equation}
The basic novel feature of the action \p{4N-gau-bose-a} as compared to the more conventional actions of supersymmetric
mechanics is the presence of the semi-dynamical spin variables $Z^k_a$ \cite{FIL08},\footnote{The
kinetic term of the variables $Z^k_a$ in the action (\ref{4N-gau-bose-1}) is
of the first-order in the time derivatives, in contrast to the dynamical variable $X_a{}^b$ with the second-order kinetic term.
Just for this reason we call $Z^k_a$, $\bar Z_k^a$ semi-dynamical variables. In the Hamiltonian (see below), they appear only in the interaction terms and enter
through the ${\rm SU}(2)$ current $S_{(ik)}$.} which has a drastic impact on the structure
of the relevant space of quantum states. These variables define extra SU(2) symmetries with the generators  \eqref{def-S}, with respect to which
the physical states carry additional spin quantum numbers and so form the appropriate SU(2) multiplets. The diagonal $su(2)$ algebra is an essential part of the ``internal'' algebra
$su(2) \subset \widehat{su}(2|1)$. Also, note the presence of the oscillator-type terms in
\p{4N-gau-bose-1} and \p{4N-gau-fermi-1}, with the intrinsic parameter $m$ as the relevant frequency.

The simplest one-particle ($n{=}1$) case of the system (\ref{4N-gau-bose-a}) was quantized in a recent paper \cite{FIS17}.
Here we consider the quantum version of the system (\ref{4N-gau-bose-a}) for an arbitrary $n$.

As shown in \cite{FI16}, at the classical level the system (\ref{4N-gau-bose-a}) describes an
${\rm SU}(2|1)\,$ supersymmetric extension of the $\mathrm{U}(2)$-spin Calogero--Moser model
\cite{GiHe,Woj,Poly1998,Poly1999,Poly2002,Poly-rev} generalizing the Calogero--Moser system  of refs. \cite{Calogero69a,Calogero69b,Calogero71,Moser,OP} to
the case with additional internal (spin) degrees of freedom.
Therefore, the basic purpose of the present paper can be formulated as a construction of new quantum multi-particle spinning Calogero--Moser type
system with deformed $\mathcal{N}{=}4, d=1$ supersymmetry.

The quantization of the Calogero-type multi-particle systems can be accomplished  by the two methods, basically leading to the same result.
One method \cite{Poly-PRL,BHV,BHKV,Poly1999,Poly1998,Poly-rev} is based on the construction of the Dunkl operators for a given system.
Using such operators  makes it possible to represent a multiparticle system as an oscillator-like system for which
the Dunkl operators play the role of generalized momentum operators.
Another way of quantizing multi-particle systems is based on considering matrix systems with additional gauge symmetries
\cite{Poly-gauge,Poly1998,Poly1999,HellRaa,Poly2001,Poly2002,Poly-rev}.
The elimination of some degrees of freedom in such matrix systems results in the standard multi-particle Calogero-type systems.
Due to the oscillator nature of matrix operators, the quantization of matrix systems is simpler
and the main task of this approach consists in finding solutions of the constraints generating gauge symmetries.
In this paper, we will mainly stick to the second method. We will present the explicit expressions of the  multi-particle operators
of deformed ${\cal N}{=}\,4$ supersymmetry, in the matrix case and for the reduced system.

The plan of the paper is as follows. In Section 2 we construct the Hamiltonian formalism for the matrix system (\ref{4N-gau-bose-a})
and show that the model indeed describes ${\rm SU}(2|1)\,$ supersymmetrization of
the ${\rm U}$(2)-spin Calogero--Moser
model \cite{GiHe,Woj,Poly1998,Poly1999,Poly2002,Poly-rev}.
In Section 3 we find, by Noether procedure, the supercharges of the underlying $\widehat{su}(2|1)$ superalgebra,
in matrix case and for a system with the reduced phase variables space.
In the latter case $\widehat{su}(2|1)$ is closed up to the constraints
generating some residual gauge invariances.
In Section 4 we construct a quantum realization of the deformed ${\cal N}{=}\,4, d=1$ superalgebra $\widehat{su}(2|1)$
for the multi-particle  Calogero--Moser system.
In the case of the reduced system with $n$ bosonic position coordinates
such a superalgebra is closed up to the generators of the $\left[\mathrm{U}(1)\right]^n$ gauge symmetry,
like in the classical case. This $\widehat{su}(2|1)$ superalgebra is represented as a sum of two $\widehat{su}(2|1)$ superalgebras.
One $\widehat{su}(2|1)$ acts in the center-of-mass sector, whereas the other operates
only on the super-variables parametrizing the quotient over  this sector. The spin operators are common for both these superalgebras.
In Sections 5 - 7 we analyze the energy spectrum in all cases: for the center-of-mass subsystem,
for the system with relative supercoordinates and in the general case, when all position operators are included.
The last Section 8 contains a Summary and outlook.

\setcounter{equation}0
\section{Hamiltonian analysis and gauge fixing}

The action (\ref{4N-gau-bose-a}) yields the canonical Hamiltonian
\be\label{Ham}
H_{\rm{total}} =H_{\rm{matrix}} -{\rm Tr}(A\,G)\,,
\ee
where
\be\label{Ham-m}
H_{\rm{matrix}} =\displaystyle \frac12\, {\rm Tr}\Big( P^2 + m^2 X^2 - 2m\bar\Psi_k \Psi^k\Big)
-\frac{S^{(ik)}S_{(ik)}}{4(X_0)^2}
+ \frac{\Psi^{(i}_0\bar\Psi^{k)}_0 S_{(ik)}}{(X_0)^2} \,.
\ee
The Hamiltonian (\ref{Ham}) involves the matrix momentum $P_a{}^b\equiv (\nabla X)_a{}^b$ and another matrix quantity
\be\label{T}
G_a{}^b\equiv  i\left[X,P \right]_a{}^b +\left\{\bar\Psi_k,\Psi^k \right\}_a{}^b
+ Z_a^k \bar Z^b_k - c\, \delta_a{}^b\,.
\ee
The action (\ref{4N-gau-bose-a}) also produces the primary constraints
\be\label{const-z}
P_{Z}{}^a_k+\frac{i}{2}\,\bar Z^a_k\approx 0\,,\qquad
P_{\bar Z}{}_a^k-\frac{i}{2}\,  Z_a^k\approx 0\,,
\ee
\be\label{const-psi}
P_{\Psi}{}_{k\,a}{}^b-\frac{i}{2}\,\bar\Psi_{k\,a}{}^b\approx 0\,,\qquad
P_{\bar\Psi}{}^k{}_{a}{}^b-\frac{i}{2}\,  \Psi^k{}_{a}{}^b\approx 0\,,
\ee
\be\label{const-A}
P_{A}{}_{a}{}^b \approx 0\,.
\ee

The constraints (\ref{const-z}), (\ref{const-psi}) are second class and so
we introduce Dirac brackets for them. As the result, we eliminate the momenta $P_{Z}{}^a_k$, $P_{\Psi}{}_{k\,a}{}^b$ and their c.c.
The residual variables obey the Dirac brackets
\be\label{Dir-br}
\left\{X_{a}{}^b, P_{c}{}^d \right\}^*=\delta_{a}^d \delta_{c}^b\,,\qquad
\left\{Z_a^k, \bar Z^b_l \right\}^*= i\delta_{l}^k \delta_{a}^b\,,\qquad
\left\{\Psi^k{}_{a}{}^b, \bar\Psi_{l\,c}{}^d \right\}^*=-i\delta_{l}^k\delta_{a}^d \delta_{c}^b\,.
\ee

Requiring the constraints (\ref{const-A}) to be preserved by the Hamiltonian (\ref{Ham}) generates secondary constraints
\be\label{const-T}
G_{a}{}^b \approx 0\,.
\ee
Despite the presence of the constant $c$ in (\ref{T}) these constraints are first class:
  with respect to the Dirac brackets (\ref{Dir-br}) they form $u(n)$ algebra,
\be\label{G-alg}
\left\{G_{a}{}^b, G_{c}{}^d \right\}^*=i\left(\delta_{a}^d G_{c}^b- \delta_{c}^b G_{a}^d \right) ,
\ee
and so produce the ${\rm U}(n)$ invariance of the action (\ref{4N-gau-bose-a})
\be\label{comp-tr-ga}
X^{\,\prime} =  e^{i\alpha}\, X\, e^{-i\alpha}\,,\quad
\Psi^{\,\prime}{}^{k} =  e^{i\alpha}\, \Psi^k\, e^{-i\alpha}\,,\quad
Z^{\prime}{}^{k} =  e^{i\alpha} Z^{k}\,, \quad A^{\,\prime} =  e^{i\alpha}\, A\,
e^{-i\alpha} - i\, e^{i\alpha} (\partial_t e^{-i\alpha}),
\ee
where $ \alpha_a{}^b(t) \in u(n) $ are $d{=}1$  gauge parameters.

In the first-order formulation, the system (\ref{4N-gau-bose-a}) is represented by the action
\be\label{4N-gau-bose-a-1st}
S_{matrix}  = \int dt \,L_{matrix}\,,
\ee
\be\label{4N-gau-bose-1st}
L_{matrix}  = {\rm Tr} \,\Big( P\dot X \Big)
+\frac{i}{2}\,{\rm Tr} \Big(\bar\Psi_k \dot\Psi^k
-\dot{\bar\Psi}_k \Psi^k \Big) + \frac{i}{2}\,
\Big(\dot {\bar Z}_k^a \, Z^k_a-\bar Z_k^a \dot Z^k_a \Big)
 -\,  H_{\rm{matrix}} +{\rm Tr}(A\,G)\,,
\ee
where $H_{\rm{matrix}}$ was defined in (\ref{Ham-m}).

Let us fix a partial gauge for the transformations (\ref{comp-tr-ga}). To this end,
we introduce the following notation for the matrix entries of $X$ and $P$:
\be\label{matrix-XP}
\begin{array}{ll}
x_a := X_a{}^a \,,\qquad & p_a:= P_a{}^a \qquad \mbox{(no summation over $a$)}\,,\\ [6pt]
x_a{}^b:= X_a{}^b \,,\qquad & p_a{}^b:= P_a{}^b  \qquad \mbox{for}\quad a\neq b\,,\\ [6pt]
x_a{}^a:= 0 \,,\qquad & p_a{}^a:= 0  \qquad\,\,\, \mbox{(no summation over $a$)}\,,
\end{array}
\ee
{\it i.e.},  $X_a{}^b =x_a \delta_a^b +x_a{}^b$, $P_a{}^b =p_a \delta_a^b +p_a{}^b$ and
${\displaystyle X_0= \frac{1}{\sqrt{n}}\,\sum_{a=1}^n x_a}$. Note that
$$\displaystyle {\rm Tr} P^2=\sum_a p_a p_a + \sum_{a\neq b}p_a{}^b p_b{}^a\,, \quad
 \displaystyle {\rm Tr} (XP)=\sum_a x_a p_a + \sum_{a\neq b}x_a{}^b p_b{}^a\,.
$$
In the notation  (\ref{matrix-XP}) the constraints (\ref{const-T}) take the form
\be\label{T-nondiag}
G_a{}^b = i(x_a-x_b)p_a{}^b - i(p_a-p_b)x_a{}^b  +i(x_a{}^c p_c{}^b-p_a{}^c x_c{}^b) +
T_a{}^b \approx 0
\ee
for $a\neq b$ and
\be\label{T-diag}
G_a{}^a = i(x_a{}^c p_c{}^a-p_a{}^c x_c{}^a) +
T_a{}^a -c\approx 0  \qquad \mbox{(no summation over $a$)}
\ee
for the diagonal elements of $G$, with
\be\label{T-def}
T_a{}^b := Z_a^k \bar Z^b_k +
\left\{\bar\Psi_k,\Psi^k \right\}_a{}^b \,.
\ee

Provided that the Calogero-like conditions $x_a \neq x_b$ are fulfilled, we can
impose the gauge
\be\label{x-fix}
x_a{}^b \approx 0\,, \qquad a\neq b\,,
\ee
for the constraints (\ref{T-nondiag}). Then we introduce Dirac brackets for the constraints  (\ref{T-nondiag}), (\ref{x-fix}) and
eliminate $x_a{}^b$ by (\ref{x-fix}) and $p_a{}^b$ by (\ref{T-nondiag}):
\be\label{p-sol}
p_a{}^b =\frac{i\,T_a{}^b}{x_a - x_b}\,, \qquad a\neq b\,.
\ee
Due to the resolved form of gauge-fixing conditions, new Dirac brackets for the remaining variables
coincide with  (\ref{Dir-br}):
\be\label{Dir-br-n}
\left\{x_{a}, p_{\,b} \right\}^{**}=\delta_{ab}\,,
\qquad
\left\{Z_a^k, \bar Z^b_l \right\}^{**}= i\delta_{l}^k \delta_{a}^b\,,\qquad
\left\{\Psi^k{}_{a}{}^b, \bar\Psi_{l\,c}{}^d \right\}^{**}=-i\delta_{l}^k\delta_{a}^d \delta_{c}^b\,.
\ee
In the gauge (\ref{x-fix}), the constraints (\ref{T-diag}) become
\be\label{T-constr}
T_a-c := T_a{}^a-c =  Z_a^k \bar Z^a_k +\left\{\bar\Psi_k,\Psi^k \right\}_a{}^a
 -c\approx 0  \qquad \mbox{(no summation over $a$)}
\ee
and they generate local  $[{\rm U}(1)]^n$ transformations of $Z_a^k$ and $\Psi^k{}_a{}^b$ with $a\neq b$.
Preservation of the conditions  (\ref{x-fix}),  $\dot x_a{}^b=\{x_a{}^b, H_{\rm{total}}\}=0\,$, allows one to express
\be\label{A-sol}
A_a{}^b =\frac{i\,p_a{}^b}{x_a - x_b}=-\frac{T_a{}^b}{(x_a - x_b)^2}\,, \qquad a\neq b\,.
\ee
Inserting (\ref{x-fix}), (\ref{T-nondiag}) and (\ref{A-sol}) into (\ref{Ham-m}), we arrive at
the reduced total Hamiltonian
\be\label{Ham-red}
H^{(red)} =H_{\rm{C-M}} -\sum_{a} A_a\Big(T_a -c\Big)\,,
\ee
where  $A_{a}=A_{a}^{a}$ (no summation over $a$) and
the generalized Calogero--Moser Hamiltonian is defined as
\be\label{Ham-CM}
\begin{array}{rcl}
H_{\rm{C-M}} &=&\displaystyle \frac12\, \sum_a \left( p_a p_a + m^2 x_a x_a\right)
+\frac12\, \sum_{a\neq b} \frac{T_a{}^b T_b{}^a}{(x_a - x_b)^2} -m {\rm Tr}\left(\bar\Psi_k \Psi^k\right)  \\ [7pt]
&& \displaystyle  -\frac{S^{(ik)}S_{(ik)}}{4(X_0)^2}
+ \frac{\Psi^{(i}_0\bar\Psi^{k)}_0 S_{(ik)}}{(X_0)^2} \,.
\end{array}
\ee

The same final result can be attained in a different way. Eliminating $A_a{}^b$, $a{\neq}b\,$, by the equations of motion
$A_a{}^b=-T_a{}^b/{(x_a-x_b)^2}$ we obtain that the action
(\ref{4N-gau-bose-a}) in the gauge (\ref{x-fix}) takes the form
\begin{eqnarray}
S_{\rm C-M} &=& \int dt \,\Bigg\{ \frac12\, \sum_{a} \Big(\dot x_a \dot x_a - m^2 x_a x_a\Big) - \frac{i}{2}\sum_{a} \Big(\bar
Z_k^a \dot Z^k_a - \dot {\bar Z}{}_k^a Z^k_a\Big)
+ \sum_{a} A_a\Big(T_a -c\Big)
\nonumber\\
&&  \qquad\quad  +\, {\rm Tr}  \Big[\frac{i}{2}\left( \bar\Psi_k \dot\Psi^k
-\dot{\bar\Psi}_k \Psi^k
\right) +m \bar\Psi_k \Psi^k\Big]  \nonumber\\
&& \qquad\quad  -\, \frac12\, \sum_{a\neq b} \frac{T_a{}^b T_b{}^a}{(x_a - x_b)^2} +\frac{S^{(ik)}S_{(ik)}}{4(X_0)^2}
-\frac{\Psi^{(i}_0\bar\Psi^{k)}_0 S_{(ik)}}{(X_0)^2}\,\Bigg\} \,.\label{act-fix}
\end{eqnarray}
The action (\ref{act-fix}) produces the Hamiltonian  (\ref{Ham-CM}), the constraints  (\ref{T-constr})  and
the brackets  (\ref{Dir-br-n}).

The important ingredients of the action (\ref{act-fix}) are
bilinear combinations of $Z^k_a$ and $\bar Z^a_k$  with the external
${\rm SU}(2)$ indices
\begin{equation}\label{S}
S_a{}_k{}^j := \bar Z^a_k Z_a^j\quad ({\rm no \ summation \ over}\ a)\,,  \qquad
S_k{}^j := \sum_a S_a{}_k{}^j \,.
\end{equation}
With respect to the Dirac brackets (\ref{Dir-br}) the objects $S_a{}_k{}^j$ for each index $a$ form $u(2)$
algebras
\begin{equation}\label{u-DB}
\left\{S_a{}_i{}^j, S_b{}_k{}^l \right\}^*  = i\delta_{ab}\left[\delta_k^j\,S_a{}_i{}^l-\delta_i^l\,S_a{}_k{}^j \right] .
\end{equation}
The object $S_k{}^j$ forms the  ``diagonal'' $u(2)$ algebra in the product of above ones
\begin{equation}\label{u-DB-0}
\left\{S_i{}^j, S_k{}^l \right\}^*  = i\left[\delta_k^j\,S_i{}^l-\delta_i^l\,S_k{}^j \right] .
\end{equation}
The triplets of the quantities (\ref{S}) (see also (\ref{def-S}))
\begin{equation}\label{S-3}
S_a^{(kj)} := \bar Z^{a(k} Z_a^{j)},\qquad
S^{(kj)} := \sum_a S_a^{(kj)}
\end{equation}
generate $su(2)$ algebras
\begin{equation}\label{su-DB}
\left\{S_a^{(ij)}, S_b^{(kl)} \right\}^*  = -i\delta_{ab}\left[\varepsilon^{ik}\,S_a^{(jl)}+\varepsilon^{jl}\,S_a^{(ik)} \right] ,
\end{equation}
\begin{equation}\label{su-DB-0}
\left\{S^{(ij)}, S^{(kl)} \right\}^*  = -i\left[\varepsilon^{ik}\,S^{(jl)}+\varepsilon^{jl}\,S^{(ik)} \right] .
\end{equation}
Below we will also use the brackets
\begin{equation}\label{su-Z}
\left\{S_{(ij)}, Z_{a}^{k} \right\}^*  = -i \delta_{(i}^{k} Z_{aj)}\,, \qquad
\left\{S^{(ij)}, \bar Z^{a}_{k} \right\}^*  = i \delta^{(i}_{k} \bar Z^{aj)} \,.
\end{equation}

One more matrix present in the action  (\ref{act-fix}) is $T_a{}^b$ defined in (\ref{T-def}).
These  quantities form $u(n)$ algebra (\ref{G-alg}) with respect to the Dirac brackets:
\be\label{T-alg}
\left\{T_{a}{}^b, T_{c}{}^d \right\}^*=i\left(\delta_{a}^d T_{c}{}^b- \delta_{c}^b T_{a}{}^d \right) .
\ee
The odd matrix variables are transformed by adjoint $u(n)$ representation:
\be\label{T-Psi}
\left\{T_{a}{}^b, \Psi^k{}_{c}{}^d \right\}^*=i\left(\delta_{a}^d \Psi^k{}_{c}{}^b - \delta_{c}^b \Psi^k{}_{a}{}^d  \right) ,
\qquad \left\{T_{a}{}^b, \Psi^k_0 \right\}^*=0\,,
\ee
\be\label{T-bPsi}
\left\{T_{a}{}^b, \bar\Psi^k{}_{c}{}^d \right\}^*=i\left(\delta_{a}^d \bar\Psi^k{}_{c}{}^b - \delta_{c}^b \bar\Psi^k{}_{a}{}^d  \right) ,
\qquad \left\{T_{a}{}^b, \bar\Psi^k_0 \right\}^*=0\,.
\ee
These $u(n)$ transformations commute with $u(2)$ transformations generated by $S_a{}_k{}^j$:
\be\label{T-S}
\left\{T_{a}{}^b, S_a{}_i{}^j\right\}^*=0\,.
\ee

Let us consider the bosonic core of the system  (\ref{act-fix}) and demonstrate that it corresponds just to the spin Calogero--Moser model.
Omitting terms with fermionic variables, we find
\begin{eqnarray}
S^{(bose)}_{\rm C-M} &=& \int dt \Bigg\{ \frac12 \sum_{a} \Big(\dot x_a \dot x_a - m^2 x_a x_a\Big) - \frac{i}{2}\sum_{a} \Big(\bar
Z_k^a \dot Z^k_a - \dot {\bar Z}{}_k^a Z^k_a\Big)
\nonumber\\
&& \qquad\qquad  + \sum_{a} A_a\Big(Z_k^a Z^k_a -c\Big)
-  \frac12\, \sum_{a\neq b} \, \frac{{\rm Tr}\left(S_a S_b\right)}{(x_a - x_b)^2} + \frac{S^{(ik)}S_{(ik)}}{4(X_0)^2}\,\Bigg\}, \label{4N-bose-fix}
\end{eqnarray}
where
\be\label{def-TrSS}
{\rm Tr}\left(S_a S_b\right):= S_a{}_k{}^j S_b{}_j{}^k
\ee
and $S_a{}_k{}^j$ are defined in (\ref{S}).
The analogous reduction of the Hamiltonian (\ref{Ham-CM}) yields
\be\label{Ham-CM-bose}
H_{\rm{C-M}} =\displaystyle \frac12\, \sum_a \left( p_a p_a + m^2 x_a x_a\right)
+\frac12\, \sum_{a\neq b} \frac{{\rm Tr}\left(S_a S_b\right)}{(x_a - x_b)^2}
-\frac{S^{(ik)}S_{(ik)}}{4(X_0)^2} \,.
\ee
The Hamiltonian (\ref{Ham-CM-bose}) contains
a potential in the center-of-mass sector with the coordinate $X_0$  (the last term in (\ref{Ham-CM-bose})). Modulo
this extra potential, the bosonic limit of the
system constructed is none other than the U(2)-spin Calogero--Moser
model which is a massive generalization of the U(2)-spin Calogero
model \cite{GiHe,Woj,Poly1999,Poly2002,Poly-rev}.
Thus the system  (\ref{act-fix}) with the Hamiltonian  (\ref{Ham-CM}) describes ${\rm SU}(2|1)$
supersymmetric extension of the U(2)-spin Calogero--Moser model.

\setcounter{equation}0
\section{Supercharges}

In this Section we will find the classical expressions for the generators of the deformed $\mathcal{N}{=}4$ supersymmetry (${\rm SU}(2|1)$ supersymmetry)
for the $n$-particle systems, both in the matrix formulation and in the case of the reduced system with $n$ position coordinates.

\subsection{Matrix system}

The odd $\mathrm{SU}(2|1)$ transformations of the component matrix fields entering (\ref{4N-gau-bose-a}) are as follows
\footnote{These transformations are a sum of the initial linear supertranslations plus extra compensating gauge transformations (\ref{tran4})
with $\lambda=2i\left(\bar\theta^+\epsilon^- -  \theta^+\bar\epsilon^- \right)A$ which are required for preserving WZ gauge for $V^{++}$.}
\be\label{143-tr}
\begin{array}{c}
\delta X =-\,\epsilon_k\Psi^k +\bar{\epsilon}^k \bar\Psi_k\,, \\ [6pt]
\displaystyle{\delta \Psi^k =i\,\bar{\epsilon}^k \left(\nabla{X}+im\,X \right)+
\frac{\bar\epsilon_j S^{(jk)}}{X_0} \,\mathbb{1} \,,\quad
\delta \bar\Psi_k =-i\, {\epsilon}_k\left(\nabla{X}-im\,X \right) + \frac{\epsilon^j S_{(jk)}}{X_0}
\,\mathbb{1}\,, }
\end{array}
\ee
whereas the supertranslations of the spin fields are represented by the ${\rm SU}(2)$ rotations
\be\label{440-tr}
\delta Z_{a}^{k} =\omega^{(kj)}Z_{a j} \,,\qquad  \delta \bar Z^{a}_{k} =-\omega_{(kj)}\bar Z^{a j}
\ee
with the composite parameters
$$
\omega^{(kj)}=\frac{\epsilon^{(k}\Psi_0^{j)}+\bar\epsilon^{(k}\bar\Psi_0^{j)}}{X_0}\,.
$$

Under the transformations (\ref{143-tr}), (\ref{440-tr}) and $\delta A=0$ the action (\ref{4N-gau-bose-a}) transforms as
\bea
&& \delta S_{matrix}= {\displaystyle \int dt \dot\Lambda_1 }\,, \nn
&& \Lambda_1 =\displaystyle -\,\frac{\epsilon_k}{2}\,{\rm Tr}\Big[\left(\nabla{X}+im\,X \right)\Psi^k\Big]
+\,\frac{\bar\epsilon^k}{2}\,{\rm Tr}\Big[\left(\nabla{X}-im\,X \right)\bar\Psi_k\Big]-
\frac{i}{2}\,\omega^{(kj)}S_{(kj)}\,.\label{total-tr}
\eea
Using (\ref{143-tr}), (\ref{440-tr}) and (\ref{total-tr}) we obtain the following expressions for Noether supercharges:
\be\label{s-charges}
\begin{array}{rcl}
{\mathcal{Q}}^k &=&\displaystyle {\rm Tr}\Big[\left(P-im\,X \right)\,\Psi^k\Big] +
\frac{i\,S^{(kj)}\,\Psi_{0\,j}}{X_0}\,,\\ [7pt]
\bar{\mathcal{Q}}_k&=& \displaystyle {\rm Tr}\Big[\left(P+im\,X \right)\bar\Psi_k\Big]-
\frac{i\,S_{(kj)}\,\bar\Psi_{0}^{j}}{X_0}\,,
\end{array}
\ee
where $P = \nabla{X}$. The generators  (\ref{s-charges}) constitute an $\widehat{su}(2|1)$ superalgebra  with respect to the Dirac brackets (\ref{Dir-br})
\be\label{Dir-br-QQ}
\left\{{\mathcal{Q}}^i, \bar{\mathcal{Q}}_k \right\}^*=-2i\, \delta^i_k\, H-2im\left(I^i_k-\,\delta^i_k\,F \right)\,,\qquad
\left\{{\mathcal{Q}}^i, {\mathcal{Q}}^k \right\}^*= 0\,,\quad
\left\{\bar {\mathcal{Q}}_i, \bar {\mathcal{Q}}_k \right\}^*=0\,.
\ee
Here, $H=H_{\rm{matrix}}$, where $H_{\rm{matrix}}$ was defined in (\ref{Ham-m}),  and also the $su(2)$ and $u(1)$
generators are present:
\be\label{def-I}
I^i_k=\varepsilon_{kj}\left[S^{(ij)}+ {\rm Tr}\Big(\Psi^{(i}\bar\Psi^{j)}\Big)\right]\,,
\ee
\be\label{def-F}
F=\frac12\,{\rm Tr}\Big(\Psi^{k}\bar\Psi_{k}\Big).
\ee
The Hamiltonian $H$ commutes with all other generators and so can be identified with the central charge operator of $\widehat{su}(2|1)$.
The rest of Dirac brackets among the generators
(\ref{Ham-m}), (\ref{s-charges}), (\ref{def-I}), (\ref{def-F}) is given by the relations
\be\label{Dir-br-allH}
\left\{H, {\mathcal{Q}}^k \right\}^*=\left\{H, \bar{\mathcal{Q}}_k \right\}^*=
\left\{H, I^i_k \right\}^*=\left\{H, F \right\}^*=0\,,
\ee
\be\label{Dir-br-allF}
\left\{F, {\mathcal{Q}}^k \right\}^*=-\frac{i}{2}\,{\mathcal{Q}}^k\,,\quad
\left\{F, \bar{\mathcal{Q}}_k \right\}^*=\frac{i}{2}\,\bar{\mathcal{Q}}_k\,,\quad
\left\{F, I^i_k \right\}^*=0\,,
\ee
\be\label{Dir-br-IQ}
\left\{I^i_k, {\mathcal{Q}}^j \right\}^*=-\frac{i}{2}\left(\delta_k^j{\mathcal{Q}}^i+\varepsilon^{ij}{\mathcal{Q}}_k\right)\,,
\quad \left\{I^i_k, \bar{\mathcal{Q}}_j \right\}^*=
\frac{i}{2}\left(\delta_j^i\bar{\mathcal{Q}}_k+\varepsilon_{kj}\bar{\mathcal{Q}}^i\right)\,,
\ee
\be\label{Dir-br-I}
\left\{I^i_k, I^j_l \right\}^*=i\left(\delta^i_lI^j_k-\delta^j_kI^i_l\right)\,.
\ee

Note that the first-order action (\ref{4N-gau-bose-a-1st}) is invariant, up to the surface term $\delta S_{matrix}= {\displaystyle \int dt \dot\Lambda_1 }$
(with the substitution $\nabla{X}=P $ in $\Lambda_1$), under the transformations (\ref{143-tr}), (\ref{440-tr}), $\delta A=0$
and
\be
\delta P=-im\left(\epsilon_k\Psi^k +\bar{\epsilon}^k \bar\Psi_k\right)-
i\,
\frac{\epsilon_k S^{(kj)}\Psi_{0\,j}+\bar\epsilon^k S_{(kj)}\bar\Psi_0^j}{X_0} \,\mathbb{1} \,.
\ee
It is worth pointing out that $\delta H=0$ and $\delta G_a{}^b=0$ under these transformations.

\subsection{Reduced system in the standard Calogero--Moser representation}

Let us compute the $\widehat{su}(2|1)$ charges for the reduced system (\ref{act-fix}) which follows from the matrix formulation
after imposing the gauge (\ref{x-fix}).

On the pattern of (\ref{matrix-XP}),
we introduce the following notation for the entries of $\Psi^k$ and $\bar\Psi_k$:
\be\label{matrix-Psi}
\begin{array}{ll}
\psi^k_a:= \Psi^k{}_a{}^a \,,\qquad & \bar\psi_a{}_k:= \bar\Psi_k{}_a{}^a \qquad \mbox{(no summation over $a$)}\,,\\ [6pt]
\psi^k{}_a{}^b:= \Psi^k{}_a{}^b \,,\qquad & \bar\psi_k{}_a{}^b:= \bar\Psi_k{}_a{}^b  \qquad \mbox{for}\quad a\neq b\,.
\end{array}
\ee
Note that
$\displaystyle {\rm Tr} (P\Psi^k)=\sum_a p_a \psi^k_a + \sum_{a\neq b} p_a{}^b \psi^k{}_b{}^a$ and
$\Psi^k_0={\displaystyle \frac{1}{\sqrt{n}}\,\sum_{a=1}^n }\psi^k_a$, $\bar\Psi^k_0={\displaystyle \frac{1}{\sqrt{n}}\,\sum_{a=1}^n }\bar\psi^k_a\,$.

In the gauge (\ref{x-fix}), supertranslations are a sum of the transformations (\ref{143-tr}), (\ref{440-tr}) and
the additional compensating gauge transformations (\ref{comp-tr-ga}) with the composite parameters
\be\label{comps-tr}
\alpha_a{}^b=i\,\frac{\epsilon_k\psi^k_a{}^b - \bar\epsilon^k\bar\psi_k{}_a{}^b}{x_a-x_b}\quad\mbox{for}\quad a\neq b\,,
\qquad\quad \alpha_a{}^b=0\quad\mbox{for}\quad a= b\,.
\ee
These  transformations preserve the conditions (\ref{x-fix}) and have the following explicit form
\be\label{mass-tr-x}
\delta x_a =-\,\epsilon_k\psi^k_a +\bar{\epsilon}^k \bar\psi_{ak}\,,
\ee
\be\label{mass-tr-psi0}
\begin{array}{rcl}
\delta \psi_a^k &=&\displaystyle{i\,\bar{\epsilon}^k \left(\dot x_a+im\,x_a \right)+
\frac{\bar\epsilon_j S^{(jk)}}{X_0} +i\sum_b \left(\alpha_a{}^b \psi^k{}_b{}^a - \psi^k{}_a{}^b\alpha_b{}^a \right) \,,}\\ [7pt]
\delta \bar\psi_{ak} &=&\displaystyle{-i\, {\epsilon}_k\left(\dot x_a-im\,x_a \right) + \frac{\epsilon^j S_{(jk)}}{X_0}
+i\sum_b \left(\alpha_a{}^b \bar\psi_{kb}{}^a - \bar\psi_{ka}{}^b\alpha_b{}^a \right)\,, }
\end{array}
\ee
\be\label{mass-tr-psi}
\begin{array}{rcl}
\delta \psi^k{}_a{}^b &=&\displaystyle{
-\frac{\bar\epsilon^k T_a{}^b}{x_a-x_b}
-i\alpha_a{}^b\left(\psi_a^k-\psi_b^k \right)+i\sum_c \left(\alpha_a{}^c \psi^k{}_c{}^b - \psi^k{}_a{}^c\alpha_c{}^b \right)\,, }\\ [7pt]
\delta \bar\psi_{ka}{}^b&=&\displaystyle{\frac{\epsilon_k  T_a{}^b}{x_a-x_b}
-i\alpha_a{}^b\left(\bar\psi_{ak}-\bar\psi_{bk} \right)
+i\sum_c \left(\alpha_a{}^c \bar\psi_{kc}{}^b - \bar\psi_{ka}{}^c\alpha_c{}^b\right)\,, }
\end{array}
\ee
\be\label{mass-tr-z}
\delta Z_{a}^{k} =\omega^{(kj)}Z_{a j} +i\sum_b \alpha_a{}^b Z_{b}^{k}\,,\qquad
\delta \bar Z^{a}_{k} =-\omega_{(kj)}\bar Z^{a j}
-i\sum_b \bar Z^{b}_{k}\alpha_b{}^a \,,
\ee
\be\label{mass-tr-A}
\delta A_{a} =i\sum_b  \frac{\alpha_a{}^b T_b{}^a + T_a{}^b \alpha_b{}^a}{x_a-x_b} \,.
\ee
An important property is that the constraints (\ref{T-constr}) are invariant with respect to these supersymmetry transformations, $\delta T_{a} =0\,$.
Also, $\delta\Big[\sum_{a} A_a\left(T_a -c\right)\Big]=0\,$.

The variation of the action (\ref{act-fix}) under the supersymmetry transformations \eqref{mass-tr-x} - \eqref{mass-tr-z} reads
$$
\delta S_{C-M}= {\displaystyle \int dt \dot\Lambda_2 }\,,
$$
where
\bea
\Lambda_2 =\displaystyle -\,\frac{\epsilon_k}{2}\,\sum_a\left(\dot{x}_a+im\,x_a \right)\psi^k_a
+\,\frac{\bar\epsilon^k}{2}\,\sum_a\left(\dot{x}_a-im\,x_a \right)\bar\psi_{ak} +\,\frac{1}{2}\,\sum_{a\neq b}\alpha_a{}^b T_b{}^a-
\frac{i}{2}\,\omega^{(kj)}S_{(kj)}\,.\label{fix-tr}
\eea
The corresponding Noether supercharges are found to be
\be\label{Q-charges}
\begin{array}{rcl}
Q^k &=&\displaystyle \sum_a\left( p_a-im x_a\right)\psi^k_a
+i\sum_{a\neq b}
\frac{T_a{}^b\,\psi^k{}_b{}^a}{x_a - x_b}
+\frac{i\,S^{(kj)}\,\Psi_{0\,j}}{X_0} \,,\\ [6pt]
\bar Q_k&=& \displaystyle  \sum_a\left( p_a+im x_a\right)\bar\psi_a{}_k
+i\sum_{a\neq b}
\frac{T_a{}^b\,\bar\psi_k{}_b{}^a}{x_a - x_b} -
\frac{i\,S_{(kj)}\,\bar\Psi_{0}^{j}}{X_0}\,,
\end{array}
\ee
where $p_a=\dot{x}_a$. These expressions can be also obtained by inserting (\ref{x-fix}), (\ref{p-sol}) into (\ref{s-charges})
and turning to the notations (\ref{matrix-XP}), (\ref{matrix-Psi}).

With respect to the Dirac brackets (\ref{Dir-br-n}), the generators  (\ref{Q-charges}) form,
up to the residual $[{\rm U}(1)]^n$ gauge transformations generated by (\ref{T-constr}), the following $\widehat{su}(2|1)$ superalgebra
\be\label{Dir-br-QbQ-red}
\begin{array}{rcl}
\left\{Q^i, \bar Q_k \right\}^{**}&=&-2i\, \delta^i_k\, H-2im\left(I^i_k-\,\delta^i_k\,F \right)+
\displaystyle{ 2i\sum_{a\neq b}
\frac{\psi^i{}_a{}^b\bar\psi_k{}_b{}^a}{(x_a - x_b)^2}\left( T_a -T_b\right) , }
\\ [8pt]
\left\{Q^i, Q^k \right\}^{**}&=& \displaystyle{ 2i\sum_{a\neq b}
\frac{\psi^i{}_a{}^b\psi^k{}_b{}^a}{(x_a - x_b)^2}\left( T_a -T_b\right)  , }
\\ [8pt]
\left\{\bar Q_i, \bar Q_k \right\}^{**}&=&\displaystyle{ 2i\sum_{a\neq b}
\frac{\bar\psi_i{}_a{}^b\bar\psi_k{}_b{}^a}{(x_a - x_b)^2}\left( T_a -T_b\right).  }
\end{array}
\ee
Here  we used that the last relation in (\ref{Dir-br-n}), being cast in the notation (\ref{matrix-XP}), (\ref{matrix-Psi}),
amounts to the relations $\left\{\psi^i_{a}, \bar\psi_{b\,k} \right\}^{**}=-i\delta_{k}^i\delta_{ab}$,
$\left\{\psi^i{}_{a}{}^b, \bar\psi_{k\,c}{}^d \right\}^{**}=-i\delta_{k}^i\delta_{a}^d \delta_{c}^b $,
In \p{Dir-br-QbQ-red}, $H=H_{\rm{C-M}}$, with $H_{\rm{C-M}}$ defined by (\ref{Ham-CM}),
and the generators $I^i_k$, $F$ were defined in (\ref{def-I}), (\ref{def-F}).

The Hamiltonian (\ref{Ham-CM}) commutes with the supercharges (\ref{Q-charges}) modulo the first-class constraints~(\ref{T-constr}):
\be\label{Dir-br-QH-red}
\left\{Q^k, H \right\}^{**}=
2\sum_{a\neq b}
\frac{T_a{}^b\psi^k{}_b{}^a}{(x_a - x_b)^3}\left( T_a -T_b\right) , \qquad
\left\{\bar Q_k, H \right\}^{**}= 2\sum_{a\neq b}
\frac{T_a{}^b\bar\psi_k{}_b{}^a}{(x_a - x_b)^3}\left(T_a -T_b\right)  .
\ee
The generators $I^i_k$, $F$ satisfy the same Dirac brackets as in (\ref{Dir-br-allH}), (\ref{Dir-br-allF}), (\ref{Dir-br-IQ}),
and (\ref{Dir-br-I}).

\setcounter{equation}0
\section{Quantum  multi-particle $\widehat{su}(2|1)$ superalgebra}

Quantum $su(2|1)$ superalgebra obtained by quantizing the Dirac brackets
(\ref{Dir-br-QQ}), (\ref{Dir-br-allH}), (\ref{Dir-br-allF}), (\ref{Dir-br-IQ}), (\ref{Dir-br-I}),
is formed by the following non-vanishing (anti)commutators :
\be\label{su21-quant}
\begin{array}{c}
{\displaystyle \left\{{\mathcal\mathbf{Q}}^i, \bar{\mathcal{\mathbf{Q}}}_k \right\}=2\, \delta^i_k\, \mathbf{H} +
2m\left(\mathbf{I}^i_k-\,\delta^i_k\,\mathbf{F} \right) ,}
\\ [6pt]
{\displaystyle \left[\mathbf{F}, {\mathcal{\mathbf{Q}}}^k \right]=\frac{1}{2}\,{\mathcal{\mathbf{Q}}}^k\,,\qquad
\left[\mathbf{F}, \bar{\mathcal{\mathbf{Q}}}_k \right]=-\frac{1}{2}\,\bar{\mathcal{\mathbf{Q}}}_k\,,}
\\ [6pt]
{\displaystyle \left[\mathbf{I}^i_k, {\mathcal{\mathbf{Q}}}^j \right]=\frac{1}{2}\left(\delta_k^j{\mathcal{\mathbf{Q}}}^i+\varepsilon^{ij}{\mathcal{\mathbf{Q}}}_k\right) ,
\qquad \left[\mathbf{I}^i_k, \bar{\mathcal{\mathbf{Q}}}_j \right]=
-\frac{1}{2}\left(\delta_j^i\bar{\mathcal{\mathbf{Q}}}_k+\varepsilon_{kj}\bar{\mathcal{\mathbf{Q}}}^i\right) ,}
\\ [6pt]
{\displaystyle \left[\mathbf{I}^i_k, \mathbf{I}^j_l \right]=\delta^j_k \mathbf{I}^i_l - \delta^i_l \mathbf{I}^j_k\,.}
\end{array}
\ee
The second- and third-order Casimir operators of $su(2|1)$ are defined by the expressions \cite{IS15}
\begin{eqnarray}\label{su21-Cas2}
{\bf C}_2&=& \Big(\frac{1}{m}\,\mathbf{H} - \mathbf{F}\Big)^2 -\frac12\, \mathbf{I}^i_k \mathbf{I}^k_i +\frac{1}{4m}\,\left[\mathbf{Q}^i,\bar{\mathbf{Q}}_i \right],
\\  [6pt]
\label{su21-Cas3}
{\bf C}_3&=& \left({\bf C}_2+\frac{1}{2}\right)\left(\frac{1}{m}\,\mathbf{H} - \mathbf{F}\right) + \frac{1}{8m}\left\lbrace\delta^j_i\left(\frac{1}{m}\,\mathbf{H} - \mathbf{F}\right)-\mathbf{I}^j_i\right\rbrace\left[\mathbf{Q}^i,\bar{\mathbf{Q}}_j\right].
\end{eqnarray}

In this Section we will present the explicit form of this deformed $\mathcal{N}{=}4$ supersymmetry algebra for multiparticle system
constructed in the previous Sections. We will do it for the matrix formulation of this system and
for the reduced system with $n$ position coordinates.

\subsection{Matrix formulation}

\subsubsection{Supercharges of the $\widehat{su}(2|1)$ superalgebra}

In the matrix formulation, the $n$-particle system is described by quantum operators ${\mathbf X}_{a}{}^b$, ${\mathbf P}_{a}{}^b$;
${\bm{\Psi}}^i{}_{a}{}^b$, $\bar{\bm{\Psi}}_{i\,a}{}^b$;
${\mathbf Z}_a^i$, $\bar{\mathbf Z}^b_i$ which satisfy the quantum counterpart of the Dirac brackets algebra (\ref{Dir-br}):
\be\label{alg-n}
\left[{\mathbf X}_{a}{}^b, {\mathbf P}_{c}{}^d \right]=i\,\delta_{a}^d \delta_{c}^b\,,\qquad
\left[{\mathbf Z}_a^k, \bar{\mathbf Z}^b_j \right]= -\,\delta_{j}^k \delta_{a}^b\,,\qquad
\left\{{\bm{\Psi}}^k{}_{a}{}^b, \bar{\bm{\Psi}}_{j\,c}{}^d \right\} = \delta_{j}^k\delta_{a}^d \delta_{c}^b\,.
\ee

The quantum supercharges are uniquely restored  by the classical expressions
(\ref{s-charges}):
\be\label{S-charges}
\begin{array}{rcl}
{\mathbf{Q}}^k &=&\displaystyle {\rm Tr}\Big[\left(\mathbf{P}-im\,{\mathbf X} \right){\bm{\Psi}}^k\Big] +
\frac{i\,{\mathbf S}^{(kj)}\,{\bm{\Psi}}_{0\,j}}{{\mathbf X}_0}\,,\\ [7pt]
\bar{\mathbf{Q}}_k&=& \displaystyle {\rm Tr}\Big[\left({\mathbf P}+im\,{\mathbf X} \right)\bar{\bm{\Psi}}_k\Big]-
\frac{i\,{\mathbf S}_{(kj)}\,\bar{\bm{\Psi}}_{0}^{j}}{{\mathbf X}_0}\,,
\end{array}
\ee
where the $su(2)$ generators are
\be\label{S-quant}
{\mathbf S}^{(ik)}=\sum_a{\mathbf Z}^{(i}_a \bar{\mathbf Z}^{k)}{}^a\,.
\ee
These generators form the quantum algebra of the corresponding diagonal external algebra \eqref{su-DB-0}.
The closure of the generators  (\ref{S-charges}) is the full $\widehat{su}(2|1)$ superalgebra (\ref{su21-quant}) with
the following even generators
\begin{eqnarray}
\label{Ham-qn}
&& {\mathbf H}={\mathbf H}^{bose}+{\mathbf H}^{fermi}\,,\\ [6pt]
&&
\label{Ham-qnb}
{\mathbf H}^{bose} = \frac12\, {\rm Tr}\left( {\mathbf P}^2 + m^2 {\mathbf X}^2 \right)
-\frac{n\,{\mathbf S}^{(ik)}{\mathbf S}_{(ik)}}{4({\mathbf X}_0)^2} \,,\\ [6pt]
&&
\label{Ham-qnf}
{\mathbf H}^{fermi} = \frac{m}{2}\, {\rm Tr}\left[{\bm{\Psi}}^k,\bar{\bm{\Psi}}_k\right] + \frac{{\bm{\Psi}}_0^{i}\bar{\bm{\Psi}}_0^{k}
{\mathbf S}_{(ik)}}{({\mathbf X}_0)^2}  \,,\\ [6pt]
\label{def-In}
&&{\mathbf I}^i_k =\varepsilon_{kj}\left[{\mathbf S}^{(ij)}+ {\rm Tr}\left({\bm{\Psi}}^{(i}\bar{\bm{\Psi}}^{j)}\right)\right]\,,\\ [6pt]
\label{def-Fn}
&& {\mathbf F} =\frac{1}{4}\,{\rm Tr}\left[{\bm{\Psi}}^{k},\bar{\bm{\Psi}}_{k}\right]\,.
\end{eqnarray}

The set of physical states of the matrix system is singled out by the $n^2$ constraints
\be\label{T-q}
{\mathbf G}_a{}^b =  \Big( i\left[{\mathbf X},{\mathbf P} \right]_a{}^b +\left\{\bar{\bm{\Psi}}_k,{\bm{\Psi}}^k \right\}_a{}^b
+ {\mathbf Z}_a^k \bar {\mathbf Z}^b_k\Big)_{\rm W}- \left(2q+1\right)\, \delta_a{}^b \simeq 0\,,
\ee
which are quantum counterparts of the classical constraints (\ref{const-T}) (the subscript ``${\rm W}$'' denotes Weyl-ordering) and should be imposed on the wave functions.
The constant $\left(2q+1\right)$ present in (\ref{T-q}) differs from the classical constant $c$ due to ordering ambiguities.
The operators  (\ref{T-q})  form $u(n)$ algebra
\be\label{alg-T-q}
\left[{\mathbf G}_a{}^b, {\mathbf G}_c{}^d \right]=  \delta_c{}^b{\mathbf G}_a{}^d -\delta_a{}^d{\mathbf G}_c{}^b\,.
\ee
It is important that all constants appearing in the diagonal part of ${\mathbf G}_a{}^b$, i.e. at $a{=}b$, are equal to $\left(2q+1\right)$.
A corollary of   (\ref{T-q}) is that $u(1)$ generator
\be\label{T-q-0}
\sum_a{\mathbf G}_a{}^a =  \sum_a{\mathbf Z}_a^k \bar {\mathbf Z}^a_k - 2nq \simeq 0
\ee
includes spin ${\mathbf Z}$-operators only.

As we will see below, the $u(n)$ constraints (\ref{T-q}) have a transparent meaning: The physical states are $su(n)$ singlets.
The constraint (\ref{T-q-0}) fixes the homogeneity degree of the physical states with respect to spin variables, whence $2q \in \mathbb{Z}_{>0}\,$.

\subsubsection{Separation of the center-of-mass sector}

Let us split the matrix quantities as
\bea
&&{\mathbf{X}}_a{}^b =\frac{1}{\sqrt{n}}\,\delta_a^b{\mathbf{X}}_0  +\hat{\mathbf{X}}_a{}^b\,,
\qquad {\mathbf{P}}_a{}^b =\frac{1}{\sqrt{n}}\,\delta_a^b{\mathbf{P}}_0  +\hat{\mathbf{P}}_a{}^b\,,\nonumber\\
&& {\bm{\Psi}}^{k}{}_{a}{}^b =\frac{1}{\sqrt{n}}\,\delta_a^b{\bm{\Psi}}^k_{0}  +\hat{\bm{\Psi}}^k{}_a{}^b\,,
\qquad \bar{\bm{\Psi}}_{k\,a}{}^b =\frac{1}{\sqrt{n}}\,\delta_a^b\bar{\bm{\Psi}}_{0\,k}  +\hat{\bar{\bm{\Psi}}}_{k\,a}{}^b\,, \label{exp-matr-0}
\eea
with
\be\label{op-matr-0}
{\mathbf{X}}_0=\frac{1}{\sqrt{n}}\sum_a{\mathbf{X}}_a{}^a\,,\quad
{\mathbf{P}}_0=\frac{1}{\sqrt{n}}\sum_a{\mathbf{P}}_a{}^a\,,\quad
{\bm{\Psi}}^k_0=\frac{1}{\sqrt{n}}\sum_a{\bm{\Psi}}^k{}_a{}^a\,,\quad
\bar{\bm{\Psi}}_{0\,k}=\frac{1}{\sqrt{n}}\sum_a\bar{\bm{\Psi}}_{k\,a}{}^a
\ee
being the center-of-mass operators and
\bea
&&\hat{\mathbf{X}}_a{}^b={\mathbf{X}}_a{}^b -\frac{1}{\sqrt{n}}\,\delta_a^b{\mathbf{X}}_0  \,,
\qquad \hat{\mathbf{P}}_a{}^b={\mathbf{P}}_a{}^b -\frac{1}{\sqrt{n}}\,\delta_a^b{\mathbf{P}}_0  \,,\nonumber\\
&& \hat{\bm{\Psi}}^k{}_a{}^b={\bm{\Psi}}^k{}_a{}^b -\frac{1}{\sqrt{n}}\,\delta_a^b{\bm{\Psi}}^k_0  \,,
\qquad \hat{\bar{\bm{\Psi}}}_{k\,a}{}^b=\bar{\bm{\Psi}}_{k\,a}{}^b -\frac{1}{\sqrt{n}}\,\delta_a^b\bar{\bm{\Psi}}_{0\,k}   \label{op-matr-tr}
\eea
the traceless parts of matrix operators.

In terms of the variables (\ref{op-matr-0}), (\ref{op-matr-tr})
the supercharges (\ref{S-charges}) are represented as
\be\label{S-matr-sum}
{\mathbf{Q}}^k ={\mathbf{Q}}_0^k+\hat{\mathbf{Q}}^k\,,\qquad
\bar{\mathbf{Q}}_k= \bar{\mathbf{Q}}_0{}_k+\hat{\bar{\mathbf{Q}}}_k\,,
\ee
where
\be\label{Q-0}
{\mathbf{Q}}_{0}^k =  \left({\mathbf P}_0-im {\mathbf X}_0\right){\bm{\Psi}}^k_0
+\frac{i\,{\mathbf S}^{(kj)}\,{\bm{\Psi}}_{0\,j}}{{\mathbf X}_0}\,,\qquad
\bar{\mathbf{Q}}_{0}{}_k = \left({\mathbf P}_0+im {\mathbf X}_0\right)\bar{\bm{\Psi}}_{0\,k} -
\frac{i\,{\mathbf S}_{(kj)}\,\bar{\bm{\Psi}}_0^{j}}{{\mathbf X}_0}
\ee
involve only the center-of-mass operators  (\ref{op-matr-0}) and spin variables, whereas
\be\label{S-matr}
\hat{\mathbf{Q}}^k =\displaystyle {\rm Tr}\Big[\left(\hat{\mathbf{P}}-im\,\hat{\mathbf X} \right)\hat{\bm{\Psi}}^k\Big] \,,\qquad
\hat{\bar{\mathbf{Q}}}_k = \displaystyle {\rm Tr}\Big[\left(\hat{\mathbf P}+im\,\hat{\mathbf X} \right)\hat{\bar{\bm{\Psi}}}_k\Big] \ee
depend on the traceless parts (\ref{op-matr-tr}).

The even operators (\ref{Ham-qn}), (\ref{Ham-qnb}), (\ref{Ham-qnf}), (\ref{def-In}), (\ref{def-Fn})
admit a similar splitting
\be\label{H-matr-sum}
{\mathbf{H}} ={\mathbf{H}}_0 +\hat{\bf{H}}\,,\qquad
{\mathbf{I}}_{k}^{i} ={\mathbf{I}}_0{}_{k}^{i} +\hat{\bf{I}}_{k}^{i}\,,\qquad
{\mathbf{F}} ={\mathbf{F}}_0 +\hat{\bf{F}}\,.
\ee
Here,
\bea
{\mathbf H}_{0} &=&\frac12 \left(\left({\mathbf P_0}\right)^2 + m^2 \left({\mathbf X_0}\right)^2\right)
+ \frac{m}{2} \left[{\bm{\Psi}}_0^k,\bar{\bm{\Psi}}_{0\,k}\right]
 - \frac{{\mathbf S}^{(ik)}{\mathbf S}_{(ik)}}{4\left({\mathbf X_0}\right)^2}+
\frac{{\mathbf S}_{(ik)}{\bm{\Psi}}_0^i \bar{\bm{\Psi}}_0^k}{\left({\mathbf X_0}\right)^2}
 \,, \label{H-tot-0}\\
{\mathbf I}_{0}{}_{k}^{i} &=&\varepsilon_{kj}\left[{\mathbf S}^{(ij)} + {\bm{\Psi}}^{(i}_0\bar{\bm{\Psi}}^{j)}_0
\right], \label{I-tot-0}\\
{\mathbf F}_{0}&=&\frac{1}{4}\,\left[{\bm{\Psi}}^{k}_0,\bar{\bm{\Psi}}_{k\,0}\right], \label{F-tot-0}
\eea
and
\begin{eqnarray}
\label{def-Hn-matr}
\hat{\bf{H}}&=& \frac12\, {\rm Tr}\left( \hat{\mathbf P}^2 + m^2 \hat{\mathbf X}^2 \right)
+ \frac{m}{2}\, {\rm Tr}\left[{\hat{\bm{\Psi}}}^k,\hat{\bar{\bm{\Psi}}}_k\right] \,,\\ [6pt]
\label{def-In-matr}
\hat{\bf{I}}^i_k&=&\varepsilon_{kj}{\rm Tr}\left(\hat{\bm{\Psi}}^{(i}\hat{\bar{\bm{\Psi}}}^{j)}\right)\,,\\ [6pt]
\label{def-Fn-matr}
\hat{\bf{F}}&=&\frac{1}{4}\,{\rm Tr}\left[\hat{\bm{\Psi}}^{k},\hat{\bar{\bm{\Psi}}}_{k}\right]\,.
\end{eqnarray}

The sets (${\mathbf{Q}}_{0}^k$, $\bar{\mathbf{Q}}_{0}{}_k$, ${\mathbf H}_{0}$, ${\mathbf I}_{0}{}_{k}^{i}$, ${\mathbf F}_{0}$)
and ($\hat{\mathbf{Q}}^k$, $\hat{\bar{\mathbf{Q}}}_k$,
$\hat{\bf{H}}$, $\hat{\bf{I}}^i_k$, $\hat{\bf{F}}$)
form $\widehat{su}(2|1)$ superalgebras (\ref{su21-quant}) on their own, with the vanishing mutual (anti)commutators:
$\left\{{\mathbf{Q}}_{0}^i,  \hat{\mathbf{Q}}^k\right\}=\left\{{\mathbf{Q}}_{0}^i,  \hat{\bar{\mathbf{Q}}}_k\right\}=0$, etc.
Thus, we have singled out the center-of-mass sector from the total system.
Note that the $\widehat{su}(2|1)$ generators $\hat{\mathbf{Q}}^k$, $\hat{\bar{\mathbf{Q}}}_k$,
$\hat{\bf{H}}$, $\hat{\bf{I}}^i_k$, $\hat{\bf{F}}$ have no action on the spin operators ${\bf Z}$ which in fact remain in the center-of-mass sector.

It is of importance that the constraints (\ref{T-q}) involve in fact only the traceless parts (\ref{op-matr-tr}) of the matrix operators (apart from the
spin variable operators). Indeed, they can be rewritten in the form
\be\label{T-q-tr}
{\mathbf G}_a{}^b = i\left[\hat{\mathbf X},\hat{\mathbf P} \right]_a{}^b +\left\{\hat{\bar{\bm{\Psi}}}_k,\hat{\bm{\Psi}}^k \right\}_a{}^b
+ {\mathbf Z}_a^k \bar {\mathbf Z}^b_k - \left(2q+n-\frac{1}{n}\right) \delta_a{}^b\simeq 0\,.
\ee
However, due to the presence of the same spin variables in the center-of-mass sector, these constraints are applicable also to the corresponding quantum
states and so accomplish a link between the two sectors.

\subsection{Quantum algebra for SU(2$|$1) spinning Calogero--Moser system}
\label{4.2}

\subsubsection{$\widehat{su}(2|1)$ superalgebra with $n$ dynamical bosons}

The quantum counterpart of the multiparticle system from Sect.\,3.2 is described by the quantum operators ${\mathbf x}_{a}$, ${\mathbf p}_{a}$;
${\bm{\psi}}^i{}_{a}$, $\bar{\bm{\psi}}_{i\,a}$; ${\bm{\psi}}^i{}_{a}{}^b$, $\bar{\bm{\psi}}_{i\,a}{}^b$, $a\neq b$;
${\mathbf Z}_a^i$, $\bar{\mathbf Z}^a_i$ which satisfy the algebra
\be\label{alg-n-gf}
\begin{array}{c}
\left[{\mathbf x}_{a}, {\mathbf p}_{b} \right]=i\,\delta_{ab}\,,\qquad
\left[{\mathbf Z}_a^k, \bar{\mathbf Z}^b_j \right]= -\,\delta_{j}^k \delta_{a}^b\,,\\ [7pt]
\left\{{\bm{\psi}}^k{}_{a}, \bar{\bm{\psi}}_{j\,b} \right\} = \delta_{j}^k\delta_{ab}\,,\qquad
\left\{{\bm{\psi}}^k{}_{a}{}^b, \bar{\bm{\psi}}_{j\,c}{}^d \right\} = \delta_{j}^k\delta_{a}^d \delta_{c}^b \;\;(a \neq b, c \neq d)\,.
\end{array}
\ee

Performing the Weyl-ordering in the quantum counterpart of (\ref{Q-charges}), we obtain the quantum supercharges:
\be\label{Q-charges-q}
\begin{array}{rcl}
    &&{\mathbf{Q}}^k = {\displaystyle\sum_a \left({\mathbf p}_a-im {\mathbf x}_a\right){\bm{\psi}}^k{}_a
 +\frac{i\,{\mathbf S}^{(kj)}\,{\bm{\Psi}}_{0\,j}}{{\mathbf X}_0}-\frac{i}{2}\sum_{a\neq b}\frac{{\bm{\psi}}^k{}_a-{\bm{\psi}}^k{}_b}{{\mathbf x}_a - {\mathbf x}_b}
+i\sum_{a\neq b}
\frac{{\mathbf T}_a{}^b\,{\bm{\psi}}^k{}_b{}^a}{{\mathbf x}_a - {\mathbf x}_b}\,,}\\
    &&\bar{\mathbf{Q}}_k = {\displaystyle\sum_a\left({\mathbf p}_a+im {\mathbf x}_a\right)\bar{\bm{\psi}}_{k\,a}-
\frac{i\,{\mathbf S}_{(kj)}\,\bar{\bm{\Psi}}_0^{j}{}}{{\mathbf X}_0}
-\frac{i}{2}\sum_{a\neq b} \frac{\bar{\bm{\psi}}_{k\,a}-\bar{\bm{\psi}}_{k\,b}}{{\mathbf x}_a - {\mathbf x}_b}
+i\sum_{a\neq b} \frac{{\mathbf T}_a{}^b\,\bar{\bm{\psi}}_k{}_b{}^a}{{\mathbf x}_a - {\mathbf x}_b}\,,}
\end{array}
\ee
where
\be\label{T-ab-q}
{\mathbf T}_a{}^b = {\mathbf Z}_a^k \bar {\mathbf Z}^b_k +
\left({\bm{\psi}}^k_a-{\bm{\psi}}^k_b\right) \bar{\bm{\psi}}_k{}_a{}^b
+ \left(\bar{\bm{\psi}}_a{}_k-\bar{\bm{\psi}}_b{}_k\right) {\bm{\psi}}^k{}_a{}^b +
\sum_{c\neq a,\,c\neq b}\left({\bm{\psi}}^k{}_a{}^c \bar{\bm{\psi}}_k{}_c{}^b + \bar{\bm{\psi}}_k{}_a{}^c{\bm{\psi}}^k{}_c{}^b \right)
\ee
are quantum counterparts of  (\ref{T-def}) at $a\neq b$
and
\be\label{0-ab-q}
{\mathbf X}_0=\frac{1}{\sqrt{n}}\,\sum_{a}{\mathbf x}_a\,,\qquad
{\bm{\Psi}}_0^i =\frac{1}{\sqrt{n}}\,\sum_{a}{\bm{\psi}}^i{}_a\,,\qquad
\bar{\bm{\Psi}}_{0\,i}=\frac{1}{\sqrt{n}}\,\sum_{a}\bar{\bm{\psi}}_{i\,a}\,.
\ee

Computing the anticommutators of the supercharges \eqref{Q-charges-q},
\begin{eqnarray}\label{A-QbQ-red}
&&\left\{{\mathbf Q}^i, \bar {\mathbf Q}_k \right\}=2\,\delta^i_k\,{\mathbf H}_n + 2m\left({\mathbf I}^i_k-\delta^i_k\,{\mathbf F}\right)-
2\sum_{a\neq b}
\frac{{\bm{\psi}}^i{}_a{}^b\bar{\bm{\psi}}_k{}_b{}^a}{\left({\mathbf x}_a - {\mathbf x}_b\right)^2}\left({\mathbf T}_a - {\mathbf T}_b\right),
\\
\label{A-QQ-red}
    &&\left\{{\mathbf Q}^i, {\mathbf Q}^k \right\}=-2\sum_{a\neq b}
\frac{{\bm{\psi}}^i{}_a{}^b{\bm{\psi}}^k{}_b{}^a}{\left({\mathbf x}_a - {\mathbf x}_b\right)^2}\left({\mathbf T}_a -{\mathbf T}_b\right),
\\
\label{A-bQbQ-red}
    &&\left\{\bar {\mathbf Q}_i, \bar {\mathbf Q}_k \right\}=-2\sum_{a\neq b}
\frac{\bar{\bm{\psi}}_i{}_a{}^b\bar{\bm{\psi}}_k{}_b{}^a}{\left({\mathbf x}_a - {\mathbf x}_b\right)^2}\left({\mathbf T}_a -{\mathbf T}_b\right)
\end{eqnarray}
we find the explicit form of the quantum even generators
\bea
{\mathbf H} &=&\frac12\,\sum_a\left({\mathbf p}_a{\mathbf p}_a + m^2 {\mathbf x}_a {\mathbf x}_a\right)
+ \frac{m}{2}\sum_a\left[{\bm{\psi}}^{k}{}_a,\bar{\bm{\psi}}_{k\,a}\right]
+ \frac{m}{2}\sum_{a\neq b}\left[{\bm{\psi}}^{k}{}_a{}^b,\bar{\bm{\psi}}_{k\,b}{}^a\right] \nonumber
\\
&&-\,\frac{{\mathbf S}^{(ik)}{\mathbf S}_{(ik)}}{4\left({\mathbf X_0}\right)^2}+
\frac{{\mathbf S}_{(ik)}{\bm{\Psi}}_0^i \bar{\bm{\Psi}}_0^k}{\left({\mathbf X_0}\right)^2}
+\frac12\,\sum_{a\neq b}\frac{{\mathbf T}_a{}^b{\mathbf T}_b{}^a}{\left({\mathbf x}_a - {\mathbf x}_b\right)^2} \,,\label{H-tot} \\
{\mathbf I}^i_k&=&\varepsilon_{kj}\left[{\mathbf S}^{(ij)} + \sum_a{\bm{\psi}}^{(i}{}_a\bar{\bm{\psi}}^{j)}{}_a
+\sum_{a\neq b} {\bm{\psi}}^{(i}{}_a{}^b \bar{\bm{\psi}}^{j)}{}_{b}{}^a \right], \label{I-tot}\\
{\mathbf F}&=&\frac{1}{4}\,\sum_a\left[{\bm{\psi}}^{k}{}_a,\bar{\bm{\psi}}_{k\,a}\right]
+ \frac{1}{4}\sum_{a\neq b}\left[{\bm{\psi}}^{k}{}_a{}^b,\bar{\bm{\psi}}_{k\,b}{}^a\right]. \label{F-tot}
\eea
The commutators of the generator ${\mathbf H}$ with odd generators ${\mathbf Q}^i$, $\bar {\mathbf Q}_i$
are a quantum generalization of (\ref{Dir-br-QH-red}). The remaining generators ${\mathbf I}^i_k$, ${\mathbf F}$
obey the same commutation relations  as in (\ref{su21-quant}).

{}From the (anti)commutators obtained we observe that the generators (\ref{Q-charges-q}), (\ref{H-tot}), (\ref{I-tot}), (\ref{F-tot}) form
the $\widehat{su}(2|1)$ superalgebra (\ref{su21-quant}) up to the differences $\left({\mathbf T}_a - {\mathbf T}_b\right)$.
However, recalling the constraints \eqref{T-constr}, this reduced system is specified also by the conditions
\be\label{T-a-q}
{\mathbf T}_a - 2q - 2\left(n-1\right) = {\mathbf Z}_a^k \bar {\mathbf Z}^a_k + \sum_{c\neq a}\left({\bm{\psi}}^k{}_a{}^c \bar{\bm{\psi}}_k{}_c{}^a
- {\bm{\psi}}^k{}_c{}^a\bar{\bm{\psi}}_k{}_a{}^c  \right)- 2q \simeq 0 \,,
\ee
which must be superimposed on the physical states. Therefore, the differences $\left({\mathbf T}_a - {\mathbf T}_b\right)$ are vanishing
on the physical states, and the physical sector of the relevant Hilbert space is closed under $\widehat{\mathrm{SU}}(2|1)$ symmetry.

It is important that the quantum constraints \eqref{T-a-q} commute with the $\widehat{su}(2|1)$ generators:
\begin{equation}\label{T-QbQ-0}
\left[{\mathbf T}_a,{\mathbf Q}^i \right]=\left[{\mathbf T}_a, \bar {\mathbf Q}_i \right]=0\,.
\end{equation}
In addition,  the quantities \eqref{T-ab-q} satisfy the algebra
\be
\left[{\mathbf T}_{a}{}^b,{\mathbf T}_{c}{}^d \right]=\delta_{c}^b\,{\mathbf T}_{a}{}^d-\delta_{a}^d\,{\mathbf T}_{c}{}^b\,,
\ee
where ${\mathbf T}_{a}{}^a={\mathbf T}_{a}$ at fixed $a$.

It is instructive to be convinced that the numerator in the last term in \p{H-tot} is indeed reduced to that for
${\rm U}(2)$ spin Calogero--Moser system \cite{Poly-rev}, when applied to the bosonic wave functions $\Phi_{bos}$ defined by the conditions
$$
\bar{\bm{\psi}}_{i\,a}\Phi_{bos} = \bar{\bm{\psi}}_{i\,a}{}^b \Phi_{bos} = 0\,.
$$
It is easy to check that in this case
$$
{\mathbf T}_a{}^b \,\Rightarrow \, {\mathbf Z}_a^k \bar {\mathbf Z}^b_k + 2(n-1)\delta^b_a
$$
and the constraint  \eqref{T-a-q} is reduced to
$$
{\mathbf Z}_a^k \bar {\mathbf Z}^a_k - 2q \simeq 0\,.
$$
Now, taking into account that in the numerator in \p{H-tot} $a\neq b$, it is easy to check that
\be
\frac12\,{\mathbf T}_a{}^b{\mathbf T}_b{}^a \,\Rightarrow \, - \frac12\,\mathbf{S}^{(ij)}_a\mathbf{S}_{b\,(ij)}+ q\left(q+1\right), \label{CoeffTT}
\qquad a \neq b\,,
\ee
where ${\mathbf S}^{(ij)}_a={\mathbf Z}_{a}^{(i} \bar{\mathbf Z}^{j)\,a}$ (no summation over $a$). The operators ${\mathbf S}^{(ij)}_a$
are just the quantum version of $S^{(ij)}_a$ defined in \p{S-3}. Foe each value of the index $a$ they generate $su(2)$ algebras and commute with each other for $a\neq b$.
The expression \p{CoeffTT} coincides with that appearing in the rational U(2) spin Calogero--Moser model, with $q$ being the pairwise spin coupling constant.\footnote{Following \cite{Poly-rev}, this model
can be referred to as the reduced matrix U(2) spin Calogero--Moser model, with $2q \in \mathbb{Z}_{>0}$. There exists another type of ${\rm U}(s)$ spin models, the so-called ``exchange-operator models''
\cite{Poly-PRL,MP92,Poly-rev,FLP2013}, for which the spin coupling constant is an arbitrary number. Our ${\rm SU}(2|1)$ supersymmetric multi-particle system yields just the first type of U(2) spin models in the bosonic sector.}

\subsubsection{Division into subsystems}

Using the simple identity
$$
\sum_{a}{\mathbf K}_a {\mathbf M}_a = \frac{1}{n}\,\sum_{a}{\mathbf K}_a \sum_{b}{\mathbf M}_b
+\frac{1}{2n}\,\sum_{a\neq b}\left({\mathbf K}_a -{\mathbf K}_b\right)\left({\mathbf M}_a -{\mathbf M}_b\right)
$$
which is valid for arbitrary $n$-vector operators ${\mathbf K}_a$, ${\mathbf M}_a$, $a=1,\ldots ,n$, and introducing
the center-of-mass quantities (\ref{0-ab-q}) and
\be\label{0P-ab-q}
{\mathbf P}_0=\frac{1}{\sqrt{n}}\,\sum_{a}{\mathbf p}_a\,,
\ee
we can represent the charges (\ref{Q-charges-q}) as the sums
\be\label{Q-sum}
{\mathbf{Q}}^k = {\mathbf{Q}}_{0}^k + {\mathbb{Q}}^k  \,,\qquad \bar{\mathbf{Q}}_k = \bar{\mathbf{Q}}_{0}{}_k+\bar{\mathbb{Q}}_k\,.
\ee
The first items ${\mathbf{Q}}_{0}^k$, $\bar{\mathbf{Q}}_{0}{}_k$ in these sums were defined in (\ref{Q-0}),
and they involve only the central-of-mass supercoordinates,
whereas the second items ${\mathbb{Q}}^k$, $\bar{\mathbb{Q}}_k$ depend only on the differences of the supercoordinates:
\be\label{Q-charges-a-b}
\begin{array}{rcl}
{\mathbb{Q}}^k &=& {\displaystyle\frac{1}{2n}\,\sum_{a\neq b}
\Big[\left({\mathbf p}_a-{\mathbf p}_b\right)-im\left({\mathbf x}_a-{\mathbf x}_b\right)\Big]
\left({\bm{\psi}}^k{}_a -{\bm{\psi}}^k{}_b\right)
}\\
&&{\displaystyle
 -\frac{i}{2}\sum_{a\neq b}\frac{{\bm{\psi}}^k{}_a-{\bm{\psi}}^k{}_b}{{\mathbf x}_a - {\mathbf x}_b}
+i\sum_{a\neq b}
\frac{{\mathbf T}_a{}^b\,{\bm{\psi}}^k{}_b{}^a}{{\mathbf x}_a - {\mathbf x}_b}\,,}\\
\bar{\mathbb{Q}}_k &=& {\displaystyle\frac{1}{2n}\,\sum_{a\neq b}
\Big[\left({\mathbf p}_a-{\mathbf p}_b\right)+im\left({\mathbf x}_a-{\mathbf x}_b\right)\Big]
\left(\bar{\bm{\psi}}_{k\,a}- \bar{\bm{\psi}}_{k\,b}\right)
}\\
&& {\displaystyle
-\frac{i}{2}\sum_{a\neq b} \frac{\bar{\bm{\psi}}_{k\,a}-\bar{\bm{\psi}}_{k\,b}}{{\mathbf x}_a - {\mathbf x}_b}
+i\sum_{a\neq b} \frac{{\mathbf T}_a{}^b\,\bar{\bm{\psi}}_k{}_b{}^a}{{\mathbf x}_a - {\mathbf x}_b}\,.}
\end{array}
\ee
Since $[{\mathbf S}^{(ij)},{\mathbf T}_a{}^b]=0$, ${\mathbf{Q}}_{0}^k$, $\bar{\mathbf{Q}}_{0}{}_k$ anticommute with
second ${\mathbb{Q}}^k$, $\bar{\mathbb{Q}}_k$:
\be\label{QQ-0}
\left\{{\mathbf{Q}}_{0}^k,  {\mathbb{Q}}^j\right\}=\left\{{\mathbf{Q}}_{0}^k,  \bar{\mathbb{Q}}_j\right\}=
\left\{\bar{\mathbf{Q}}_{0}{}_k,  {\mathbb{Q}}^j\right\}=\left\{\bar{\mathbf{Q}}_{0}{}_k,  \bar{\mathbb{Q}}_j\right\}=0   \,.
\ee
This implies that the bosonic generators (\ref{H-tot}), (\ref{I-tot}), (\ref{F-tot}) can also be represented as similar sums,
\be\label{B-sum}
{\mathbf{H}} = {\mathbf{H}}_{0} + {\mathbb{H}} \,,\qquad
{\mathbf{I}}^i_k = {\mathbf{I}}_{0}{}^i_k + {\mathbb{I}}^i_k \,,\qquad
{\mathbf{F}} = {\mathbf{F}}_{0} + {\mathbb{F}},
\ee
where ${\mathbf{H}}_{0}, {\mathbf{I}}_{0}{}^i_k $ and ${\mathbf{F}}_{0}$ are given by eqs. (\ref{H-tot-0}), (\ref{I-tot-0}), (\ref{F-tot-0}) and so
involve only the center-of-mass coordinates, while the rest of operators is defined by the expressions
\bea
{\mathbb{H}} &=&\frac{1}{4n}\,\sum_{a\neq b}
\left(\left({\mathbf p}_a-{\mathbf p}_b\right)^2 + m^2 \left({\mathbf x}_a -{\mathbf x}_b\right)^2\right)
+\frac12\,\sum_{a\neq b}\frac{{\mathbf T}_a{}^b{\mathbf T}_b{}^a}{\left({\mathbf x}_a - {\mathbf x}_b\right)^2}  \nonumber
\\
&&
+ \frac{m}{4n}\,\sum_{a\neq b}\left[\left({\bm{\psi}}^{k}{}_a -
{\bm{\psi}}^{k}{}_b\right),\left(\bar{\bm{\psi}}_{k\,a}-\bar{\bm{\psi}}_{k\,b}\right)\right]
+ \frac{m}{2}\sum_{a\neq b}\left[{\bm{\psi}}^{k}{}_a{}^b,\bar{\bm{\psi}}_{k\,b}{}^a\right] \,, \label{H-tot-ab}\\
{\mathbb{I}}^i_k&=&\varepsilon_{kj}\left[\frac{1}{2n}\,\sum_{a\neq b}\left({\bm{\psi}}^{(i}{}_a-{\bm{\psi}}^{(i}{}_b\right)
\left(\bar{\bm{\psi}}^{j)}{}_a - \bar{\bm{\psi}}^{j)}{}_b \right)
+\sum_{a\neq b} {\bm{\psi}}^{(i}{}_a{}^b \bar{\bm{\psi}}^{j)}{}_{b}{}^a \right], \label{I-tot-ab}\\
{\mathbb{F}}&=&\frac{1}{8n}\,\sum_{a\neq b}\left[\left({\bm{\psi}}^{k}{}_a -
{\bm{\psi}}^{k}{}_b\right),\left(\bar{\bm{\psi}}_{k\,a}-\bar{\bm{\psi}}_{k\,b}\right)\right]
+ \frac{1}{4}\sum_{a\neq b}\left[{\bm{\psi}}^{k}{}_a{}^b,\bar{\bm{\psi}}_{k\,b}{}^a\right]. \label{F-tot-ab}
\eea

The sets of the generators (${\mathbf{Q}}_{0}^k$, $\bar{\mathbf{Q}}_{0}{}_k$, ${\mathbf{H}}_{0}$, ${\mathbf{I}}_{0}{}^i_k$, ${\mathbf{F}}_{0}$)
and (${\mathbb{Q}}^k$, $\bar{\mathbb{Q}}_k$, ${\mathbb{H}}$, ${\mathbb{I}}^i_k$, ${\mathbb{F}}$)
form two separate mutually (anti)commuting  $\widehat{su}(2|1)$ superalgebras. Note that second set generates an $\widehat{su}(2|1)$ superalgebra up to the constraints,
as in \eqref{A-QbQ-red}, \eqref{A-QQ-red}, \eqref{A-bQbQ-red}. Also, note that the ``internal'' SU(2) generators \eqref{I-tot-ab} (appearing in the  anticommutator of supercharges)
act on the indices $i,j$ of the fermionic operators ${\bm{\psi}}^i{}_{a}-{\bm{\psi}}^i{}_{b}$, $\bar{\bm{\psi}}_{i\,a}-\bar{\bm{\psi}}_{i\,b}$,
${\bm{\psi}}^i{}_{a}{}^b$, $\bar{\bm{\psi}}_{i\,a}{}^b$, $a\neq b$\,,
while the indices $i,j$ of the spin operators ${\mathbf Z}_a^i$, $\bar{\mathbf Z}^a_i$ are subject to the action of the external SU(2) generators \eqref{S-quant}.

\subsection{Subsystems of $\mathcal{N}{=}4$ supersymmetric Calogero--Moser model}

We have found that the $\mathcal{N}{=}4$ supersymmetric $n$-particle Calogero--Moser system
is a direct sum of two subsystems with different realizations of the $\widehat{su}(2|1)$ generators.

The generators (${\mathbf{Q}}_{0}^k$, $\bar{\mathbf{Q}}_{0}{}_k$, ${\mathbf{H}}_{0}$, ${\mathbf{I}}_{0}{}^i_k$, ${\mathbf{F}}_{0}$)
act in the sector of the center-of-mass operators
(${\mathbf X_0}$, ${\mathbf P_0}$, ${\bm{\Psi}}_0^i$, $\bar{\bm{\Psi}}_{0\,i}$)
and the spin operators (${\mathbf Z}_a^i$, $\bar{\mathbf Z}^a_i$).
The second set of the $\widehat{su}(2|1)$ generators ($\hat{\mathbf{Q}}^k$, $\hat{\bar{\mathbf{Q}}}_k$,
$\hat{\bf{H}}$, $\hat{\bf{I}}^i_k$, $\hat{\bf{F}}$)
act, in the matrix formulation, within the sector of the traceless operators ($\hat{\mathbf X}$, $\hat{\mathbf P}$,
$\hat{\bm{\Psi}}^i$, $\hat{\bar{\bm{\Psi}}}_{i}$).
Physical states in this subsystem are specified also by the spin operators (${\mathbf Z}_a^i$, $\bar{\mathbf Z}^a_i$)
which are present in the $u(n)$ constraints (\ref{T-q-tr}). These constraints also specify physical states in the center-of-mass  sector involving the same spin operators.
In the reduced formulation, the generators (${\mathbb{Q}}^k$, $\bar{\mathbb{Q}}_k$, ${\mathbb{H}}$, ${\mathbb{I}}^i_k$, ${\mathbb{F}}$)
are spanned by the set of operators
(${\mathbf x}_{a}-{\mathbf x}_{b}$, ${\mathbf p}_{a}-{\mathbf p}_{b}$,
${\bm{\psi}}^i{}_{a}-{\bm{\psi}}^i{}_{b}$, $\bar{\bm{\psi}}_{i\,a}-\bar{\bm{\psi}}_{i\,b}$,
${\bm{\psi}}^i{}_{a}{}^b$, $\bar{\bm{\psi}}_{i\,a}{}^b$, $a\neq b$;
${\mathbf Z}_a^i$, $\bar{\mathbf Z}^a_i$).
It should be pointed out that the spin operators ${\mathbf Z}_a^i$, $\bar{\mathbf Z}^a_i$ have a non-zero action on the physical states with $q\,{\neq}\,0$ for all
subsystems defined above and listed below.

The just described structure of the considered system suggests that we can consider three subsystems:
\begin{description}
\item[I)\,\,\,\,\,\,]
The center-of-mass sector spanned by the quantum operators (${\mathbf X_0}$, ${\mathbf P_0}$, ${\bm{\Psi}}_0^i$, $\bar{\bm{\Psi}}_{0\,i}$, ${\mathbf Z}_a^i$, $\bar{\mathbf Z}^a_i$)
and the symmetry operators (${\mathbf{Q}}_{0}^k$, $\bar{\mathbf{Q}}_{0}{}_k$, ${\mathbf{H}}_{0}$, ${\mathbf{I}}_{0}{}^i_k$, ${\mathbf{F}}_{0}$);
\item[II)\,\,\,\,]
The pure Calogero--Moser multi-particle sector with the center-of-mass sector separated. It is spanned by the quantum operators
($\hat{\mathbf X}$, $\hat{\mathbf P}$,
$\hat{\bm{\Psi}}^i$, $\hat{\bar{\bm{\Psi}}}_{i}$) in the matrix formulation or by (${\mathbf x}_{a}-{\mathbf x}_{b}$, ${\mathbf p}_{a}-{\mathbf p}_{b}$,
${\bm{\psi}}^i{}_{a}-{\bm{\psi}}^i{}_{b}$, $\bar{\bm{\psi}}_{i\,a}-\bar{\bm{\psi}}_{i\,b}$,
${\bm{\psi}}^i{}_{a}{}^b$, $\bar{\bm{\psi}}_{i\,a}{}^b$, $a\neq b$) in the reduced formulation. In both formulations, this subsystem also involves the spin operators
${\mathbf Z}_a^i$, $\bar{\mathbf Z}^a_i$. The ${\rm SU}(2|1)$ symmetry generators are
($\hat{\mathbf{Q}}^k$, $\hat{\bar{\mathbf{Q}}}_k$,
$\hat{\bf{H}}$, $\hat{\bf{I}}^i_k$, $\hat{\bf{F}}$) or (${\mathbb{Q}}^k$, $\bar{\mathbb{Q}}_k$, ${\mathbb{H}}$, ${\mathbb{I}}^i_k$, ${\mathbb{F}}$);
\item[III)\,]
The full Calogero--Moser multi-particle system which contains the center-of-mass sector and so is spanned by the set of all
quantum operators.  The $\widehat{{\rm SU}}(2|1)$ symmetry generators are sums of the $\widehat{\mathrm{SU}}(2|1)$ generators acting in the two previously
defined  sectors.
\end{description}

Now we are prepared to  determine the energy spectrum of all these systems.

\setcounter{equation}0
\section{Center-of-mass subsystem with $n$ sets of spin variables}

In this section we consider the subsystem {\bf I)} which describes the center-of-mass sector with the Hamiltonian
${\mathbf{H}}_0$ (\ref{H-tot-0}).

The center-of-mass supercoordinates (\ref{0-ab-q}), (\ref{0P-ab-q}) satisfy the following (anti)commutation relations
\be \label{0-q-al}
\left[{\mathbf X_0}, {\mathbf P_0} \right]=i\,,\qquad
\left\{{\bm{\Psi}}_0^i, \bar{\bm{\Psi}}_{0\,k} \right\} = \delta^{i}_k \,,
\ee
while those for the spin variables read
\be \label{z-q-al}
\left[{\mathbf Z}_a^i, \bar{\mathbf Z}^b_k \right]= -\,\delta_{k}^i \delta_{a}^b \,.
\ee
We will use the following realization of the operator relations (\ref{0-q-al}), (\ref{z-q-al})
\be\label{realiz-x0}
{\mathbf X}_0=x_0\,,\quad {\mathbf P}_0=-i\frac{\partial}{\partial x_0}\,,\qquad
{\bm{\Psi}}^i_{0}={\psi}^i_{0} \,,\quad  \bar{\bm{\Psi}}_{0\,i} = \frac{\partial}{\partial {\psi}^i_{0}}\,,
\ee
\be\label{realiz-z}
{\mathbf Z}^i_a = z^i_a\,,\quad \bar{\mathbf Z}_i^a= \frac{\partial}{\partial z^i_a} \,,
\ee
where $x_0$ is a real commuting variable, $z^i_a$ are complex commuting variables and ${\psi}^i_{0}$ are complex Grassmann variables.
In this realization the Hamiltonian (\ref{H-tot-0}) takes the form
\be \label{H-quant-0}
{\mathbf H}_{0} =\frac12 \left(-\frac{\partial^2}{\partial x_0{}^2} + m^2 x_0{}^2 \right)
+ m \left({\psi}^i_{0}\,\frac{\partial}{\partial {\psi}^i_{0}} -1 \right)
 +\frac{1}{x_0{}^2} \left( -\frac{1}{4}\,{\mathbf S}^{(ik)}{\mathbf S}_{ik}+
{\mathbf S}^{(ik)}{\psi}_{0\,i}\,\frac{\partial}{\partial {\psi}^k_{0}} \right)
 \,,
\ee
where ${\displaystyle {\mathbf S}_{(ij)}=\sum_a z_{a\,(i} \frac{\partial}{\partial z_a^{j)}} }\,$.
Wave function $\Phi^{(2q)}(x_0,z^i_a, {\psi}^i_{0})$ is subject to the $n$ constraints originating from \eqref{T-q-tr}:
\be \label{constr-wf-0n}
    {\mathbf G}_a{}^a\,\Phi^{(2q)} = \left(z_a^k \frac{\partial}{\partial z_a^k} -2q\right)\Phi^{(2q)}=0\,,\qquad
a=1,\ldots,n\,.
\ee
The solution of eqs. \eqref{constr-wf-0n} is a function which is homogeneous of degree $2q$ with respect to each set of spin variables $z_a^k\,$.
So the number $q$ taking positive integer and half-integer values can be treated as a spin associated with every SU(2) group generated by the quantum generators
\bea\label{S_a}
    {\mathbf S}^{(ij)}_a=z_{a\,(i} \frac{\partial}{\partial z_a^{j)}}\,,\qquad \mbox{(no summation over $a$)}.
\eea
The number $s$ will be associated with the diagonal SU(2) group generated by
\bea
    {\mathbf S}^{(ij)}=\sum_{a=1}^n{\mathbf S}^{(ij)}_a,
\eea
and, in what follows, will be referred to as ``${\rm SU}(2)$ spin s''.
Since $\Phi^{(2q)}$ is transformed in the direct product of $n$ spin $q$ ${\rm SU}(2)$ representations, the maximal external ${\rm SU}(2)$ spin is just $s = nq$.
It will be convenient to expand $\Phi^{(2q)}$ into irreducible multiplets of the diagonal ${\rm SU}(2)$, with spins running in the intervals $0,1\ldots nq$ (for $2nq$ even) or $1/2,3/2\ldots nq$ (for $2nq$ odd).

As an illustration, we dwell on two lower-$n$ cases.

\vspace{0.5cm}

\centerline{\underline{$n=2$}}

\vspace{0.3cm}

In this case the wave function $\Phi_0(x_0,z^i_1, z^i_2, {\psi}^i_{0})$ is subject to two constraints
\bea \label{constr-wf-02}
\left({\mathbf T}_1-2q\right)\Phi_0^{(2q)} = \left(z_1^k \frac{\partial}{\partial z_1^k} -2q\right)\Phi_0^{(2q)}=0\,,\;
\left({\mathbf T}_2-2q\right)\Phi_0^{(2q)} = \left(z_2^k \frac{\partial}{\partial z_2^k} -2q\right)\Phi_0^{(2q)}=0\,.
\eea
Their general solution is
\be\label{expand-wf-01}
\Phi_0^{(2q)} = \left(z_1 z_2\right)^{2q} \phi +\sum_{s=1}^{2q}\left( z_1 z_2\right)^{2q-s}
z_1^{i_1}\ldots z_1^{i_{s}}\,z_2^{i_{s+1}}\ldots z_2^{i_{2s}}\,\Phi_{(i_1 \ldots i_{2s})}\,,
\ee
where $\left(z_1 z_2\right):=z_1^i z_{2i}$. The component wave functions $\phi$, $\Phi_{(i_1 \ldots i_{2s})}$
in the expansion (\ref{expand-wf-01})are functions of $x_0$, ${\psi}^i_{0}$. They form irreducible ${\rm SU}(2)$ multiplets with spins $s = 0, 1,\ldots 2q\,$.
Their expansions with respect to ${\psi}^i_{0}$ are
\bea\label{expand-wf-001}
\phi &=& a_{+}+{\psi}^i_{0}b_i+\left({\psi}_{0}\right)^2 a_{-}\,,\\
\Phi_{(i_1 \ldots i_{2s})} &=& A_{+(i_1 \ldots i_{2s})}
+{\psi}_{0\,(i_1}B_{i_2 \ldots i_{2s})}  + {\psi}^j_{0}C_{j(i_1 \ldots i_{2s})}
+\left({\psi}_{0}\right)^2 A_{-(i_1 \ldots i_{2s})}\,. \label{expand-wf-011}
\eea
All components in these expansions are functions of $x_0$ only.
In the bosonic wave function $\Phi_0^{(2q)}$, the fields $a_{\pm}$, $A_{\pm(i_1 \ldots i_{2s})}$
are bosonic, whereas $b_i$, $B_{(i_2\ldots i_{2s})}$, $C_{(ji_1 \ldots i_{2s})}$ are fermionic. The ${\rm SU}(2)$ spins of the component
wave functions are counted with respect to the ``internal'' ${\rm SU}(2)$ with the generators \p{I-tot-0} which contain, besides the part acting on the bosonic spin variables, also
the one acting on the fermionic variables.

Let us determine the eigenvalues of the Hamiltonian (\ref{H-quant-0}) on the wave function (\ref{expand-wf-01}),
{\cal i.e.} solve the stationary Schr\"{o}dinger equation
\be\label{Schrod-0}
{\mathbf H}_{0}\,\Phi_0^{(2q,\ell)}=E_{s,\ell}\,\Phi_0^{(2q,\ell)} \,.
\ee
As a prerequisite, we adduce the following eigenvalue relations
\be \label{eq-1-2}
-\frac12\,{\mathbf S}^{(ij)}{\mathbf S}_{(ij)}\,
z_1^{k_1}\ldots z{}^{k_{p}}_1 z_2^{k_{p+1}}\ldots z{}^{k_{2s}}_2 A_{(k_1\ldots k_{2s})}=  s\left(s+1 \right)
z_1^{k_1}\ldots z{}^{k_{p}}_1 z_2^{k_{p+1}}\ldots z{}^{k_{2s}}_2 A_{(k_1\ldots k_{2s})}\,,
\ee
\be \label{eq-2-2}
{\mathbf S}_{(ij)}\,
(z_1 z_{2})= 0\,,\qquad
{\mathbf S}^{(ij)}{\mathbf S}_{(ij)}\, {\psi}^k_{0}z{}_{a\, k}=
- {\mathbf S}^{(ij)}{\psi}_{0\,i}\,\frac{\partial}{\partial {\psi}^j_{0}}\, {\psi}^k_{0}z{}_{a\, k}=
-\frac{3}{2}\, {\psi}^k_{0}z{}_{a\, k}\,,
\ee
\be \label{eq-3-2}
\begin{array}{l}
{\displaystyle  {\mathbf S}^{(ij)}{\psi}_{0\,i}\,\frac{\partial}{\partial {\psi}^j_{0}}\,
{\psi}^n_{0}z_1^{k_1}\ldots z{}^{k_{p}}_1 z_2^{k_{p+1}}\ldots z{}^{k_{2s}}_2 A_{(nk_1\ldots k_{2s})}= } \qquad\qquad\qquad\qquad\qquad\qquad\\ [8pt]
\qquad\qquad\qquad \qquad\qquad\qquad\qquad {\displaystyle = - s\,
{\psi}^n_{0}z_1^{k_1}\ldots z{}^{k_{p}}_1 z_2^{k_{p+1}}\ldots z{}^{k_{2s}}_2 A_{(nk_1\ldots k_{2s})}\,. }
\end{array}
\ee
\be \label{eq-4-2}
\begin{array}{l}
{\displaystyle  {\mathbf S}^{(ij)}{\psi}_{0\,i}\,\frac{\partial}{\partial {\psi}^j_{0}}\,
z_1^{k_1}\ldots z{}^{k_{p}}_1 z_2^{k_{p+1}}\ldots z{}^{k_{2s}}_2 {\psi}_{0\,(k_1}A_{k_2\ldots k_{2s})}= } \quad\qquad\qquad\qquad\qquad\qquad\\ [8pt]
\qquad\qquad\qquad \qquad\qquad\qquad\qquad {\displaystyle = (s+1)\,
z_1^{k_1}\ldots z{}^{k_{p}}_1 z_2^{k_{p+1}}\ldots z{}^{k_{2s}}_2 {\psi}_{0\,(k_1}A_{k_2\ldots k_{2s})} }.
\end{array}
\ee
They can be easily checked and shown to be valid for an arbitrary $p\leq 2s$.
Due to these relations, all the component fields in the expansions (\ref{expand-wf-001}), (\ref{expand-wf-011})
of the wave function (\ref{expand-wf-01}) are eigenstates of the center-of-mass Hamiltonian (\ref{H-quant-0}) with the spin $s$ of the diagonal SU(2) group given by ${\mathbf S}^{(ij)}$.
The equation (\ref{Schrod-0}) amounts to the following equations for the component wave functions
\bea\label{eq-wf-1}
    &&\frac12 \left[ -\frac{\partial^2}{\partial x_0^{\;2}} + m^2 x_0{}^2\right] a^{(\ell)}_{\pm}=\left(E_{0,\ell}\pm m \right) a^{(\ell)}_{\pm}\,,\nn
    &&\frac12 \left[ -\frac{\partial^2}{\partial x_0^{\;2}} + m^2 x_0{}^2\right] b^{(\ell)}_{i}= E_{0,\ell}\,b^{(\ell)}_{i}\,,
\eea
\bea\label{eq-wf-2}
    &&\frac12 \left[ -\frac{\partial^2}{\partial x_0^{\;2}} + m^2 x_0{}^2
+\frac{s(s+1)}{x_0{}^2}\right] A^{(\ell)}_{\pm(i_1 \ldots {i_{2s}})}=
\left(E_{s,\ell}\pm m \right) A^{(\ell)}_{\pm(i_1 \ldots {i_{2s}})}\,,\nn
    &&\frac12 \left[-\frac{\partial^2}{\partial x_0^{\;2}} + m^2 x_0{}^2
+\frac{(s+1)(s+2)}{x_0{}^2}\right] B^{(\ell)}_{(i_1 \ldots {i_{2s-1}})}= E_{s,\ell}\, B^{(\ell)}_{(i_1 \ldots {i_{2s-1}})}\,,\nn
    &&\frac12 \left[-\frac{\partial^2}{\partial x_0^{\;2}}+ m^2 x_0{}^2+\frac{s(s-1)}{x_0{}^2}\right]
C^{(\ell)}_{(i_1 \ldots {i_{2s+1}})}= E_{s,\ell}\, C^{(\ell)}_{(i_1 \ldots {i_{2s+1}})}
\,,
\eea
where \p{eq-wf-1} corresponds to $s=0$, while in \p{eq-wf-2} $s$ runs over $2q \geq 1$ values, $s=1,\ldots,2q$\,.

The equations  \eqref{eq-wf-1} for the fields $a^{(\ell)}_{\pm}$ and $b^{(\ell)}_i$ have the form
\be\label{osc-wf-2}
    \frac12 \left[-\frac{\partial^2}{\partial x_0^{\;2}} + m^2 x_0{}^2\right]f^{(\ell)}(x_0)=\mathscr{E}_\ell\,f^{(\ell)}(x_0)
\ee
and describe the excitations of oscillators.
The standard solutions of the equation \eqref{osc-wf-2} are given via Hermite polynomials ${H}_{\ell}$ as
\be
f^{(\ell)}(x_0)=\mbox{H}_{\ell}\left(x_0\right)\exp{\left(-mx_0{}^2/2\right)}\,,\quad \ell=0,1,2,\ldots\,,
\ee
and have the energies
\be
\mathscr{E}_\ell = m\left(\ell+1/2\right).
\ee
Thus, the energy spectrum reads
\be
    E_{0,\ell} = m\left(\ell - \frac{1}{2}\right).
\ee
It is worth pointing out that the solution for ${\cal N}{=}\,4$ supersymmetric harmonic oscillator \eqref{eq-wf-1}
was originally given in \cite{Sm}.

The equations \eqref{eq-wf-2} for the fields $A_{\pm(i_1 \ldots {i_{2s}})}$, $B_{(i_1 \ldots {i_{2s-1}})}$ and $C_{(i_1 \ldots {i_{2s+1}})}$
have the generic form
\be\label{eq-conf}
\frac12 \left[-\frac{\partial^2}{\partial x_0^{\;2}}+ m^2 x_0{}^2
+\frac{\gamma\left(\gamma-1 \right)}{x_0{}^2}\right] f^{(\ell)}(x_0)=
\mathscr{E}_\ell^\prime \, f^{(\ell)}(x_0)\,,
\ee
where $\gamma$ is a constant.  It is the  well-known equation describing quantum states of non-relativistic particle moving in a sum of the one-dimensional oscillator and
conformal inverse-square potentials,  and it has
the following general solution  (see, e.g., \cite{Calogero69a,Calogero71,Per})
\be
f^{(\ell)}(x_0)=\sqrt{\frac{2\ell!}{\Gamma(\ell+\gamma+1/2)}}\,
x_0{}^\gamma\,L^{(\gamma - 1/2)}_\ell(mx_0{}^2)\,\exp(-mx_0{}^2/2)\,,\quad \ell=0,1,2,\ldots\,,\label{eq-conf-sol}
\ee
where $L^{(\gamma - 1/2)}_\ell$ is a generalized Laguerre polynomial.
The corresponding energy levels are
\be
\mathscr{E}_{\gamma ,\ell}^\prime=
m\left(2\ell+\gamma+\frac12 \right). \label{eq-conf-ener}
\ee

The general solution  \eqref{eq-conf-sol} was used in ref. \cite{FIS17}
to reveal the energy spectrum of the one-particle system with one set of the spin variables.
Each equation in the set \eqref{eq-wf-2} has the form of \eqref{eq-conf},  the parameter $\gamma$ being $s+1$, $s+2$ and $s$, respectively.
Thus, the energy of the states of spins $s$ and $s+1/2$ described by the wave functions $A_{+(i_1 \ldots {i_{2s}})}$ and $C_{(i_1 \ldots {i_{2s+1}})}$,
is equal to
$$
    E_{s,\ell}=m\left( 2\ell+s+1/2\right),\quad \ell=0,1,2,\ldots\,,\qquad s=1,\ldots,2q\,.
$$
The energy of the states $A_{-(i_1 \ldots {i_{2s}})}$ of spin $s$
and the states  $B_{(i_1 \ldots {i_{2s-1}})}$ of spin $s+1/2$ is given only for excited states by the same expression
$$
    E_{s,\ell}=m\left( 2\ell+s+1/2\right),\quad \ell=1,2,3,\ldots\,,\qquad s=1,\ldots,2q\,.
$$
The lowest energy for these states corresponds to  $s=1$, $\ell=0$ and equals
\be\label{vac-en-2}
E_{\rm min}=\frac{3m}{2}\,.
\ee
At $q=1/2$ we have the picture drawn in the Figure 1.

\begin{figure}[ht]
\begin{center}
\begin{picture}(460,230)
\put(20,5){\line(0,1){210}}
\put(20,215){\vector(0,1){10}}

\put(195,5){\line(0,1){10}}
\put(195,25){\line(0,1){10}}
\put(195,45){\line(0,1){10}}
\put(195,65){\line(0,1){10}}
\put(195,85){\line(0,1){10}}
\put(195,105){\line(0,1){10}}
\put(195,125){\line(0,1){10}}
\put(195,145){\line(0,1){10}}
\put(195,165){\line(0,1){10}}
\put(195,185){\line(0,1){10}}
\put(195,205){\line(0,1){10}}

\put(25,20){\line(1,0){430}}
\put(25,50){\line(1,0){430}}
\put(25,80){\line(1,0){430}}
\put(25,110){\line(1,0){430}}
\put(25,140){\line(1,0){430}}
\put(25,170){\line(1,0){430}}
\put(25,200){\line(1,0){430}}

\multiput(230,80)(15,0){3}{\circle*{8}}
\multiput(230,140)(15,0){3}{\circle*{8}}
\multiput(230,200)(15,0){3}{\circle*{8}}
\multiput(280,136)(15,0){2}{{\large $\times$}}
\multiput(280,196)(15,0){2}{{\large $\times$}}
\multiput(325,76)(15,0){4}{{\large  $\times$}}
\multiput(325,136)(15,0){4}{{\large  $\times$}}
\multiput(325,196)(15,0){4}{{\large  $\times$}}
\multiput(405,140)(15,0){3}{\circle*{8}}
\multiput(405,200)(15,0){3}{\circle*{8}}
\put(60,20){\circle*{8}}
\put(60,50){\circle*{8}}
\put(60,80){\circle*{8}}
\put(60,110){\circle*{8}}
\put(60,140){\circle*{8}}
\put(60,170){\circle*{8}}
\put(60,200){\circle*{8}}
\multiput(96,46)(15,0){2}{{\large $\times$}}
\multiput(96,76)(15,0){2}{{\large $\times$}}
\multiput(96,106)(15,0){2}{{\large $\times$}}
\multiput(96,136)(15,0){2}{{\large $\times$}}
\multiput(96,166)(15,0){2}{{\large $\times$}}
\multiput(96,196)(15,0){2}{{\large $\times$}}
\put(160,80){\circle*{8}}
\put(160,110){\circle*{8}}
\put(160,140){\circle*{8}}
\put(160,170){\circle*{8}}
\put(160,200){\circle*{8}}

\put(-5,16){{\small ${\displaystyle -\frac{m}{2}}$}}
\put(4,46){{\small ${\displaystyle\frac{m}{2}}$}}
\put(0,76){{\small ${\displaystyle\frac{3m}{2}}$}}
\put(0,106){{\small ${\displaystyle\frac{5m}{2}}$}}
\put(0,136){{\small ${\displaystyle\frac{7m}{2}}$}}
\put(0,166){{\small ${\displaystyle\frac{9m}{2}}$}}
\put(-5,196){{\small ${\displaystyle\frac{11m}{2}}$}}

\put(0,225){{\small ${\mathbf H}_0$}}
\put(235,0){\small $A_{+(ij)}$}
\put(290,0){\small $B_i$}
\put(350,0){\small $C_{(ijk)}$}
\put(410,0){\small $A_{-(ij)}$}
\put(55,0){\small $a_{+}$}
\put(105,0){\small $b_i$}
\put(160,0){\small $a_{-}$}
\end{picture}
\end{center}
\caption{The degeneracy of energy levels of ${\mathbf H}_0$ for $n\,{=}\,2$ and $q\,{=}\,1/2$\,. Circles and crosses represent bosonic and fermionic states,
respectively. On the left from the dotted vertical line the degeneracy corresponding to harmonic oscillator \cite{Sm, IS14a} is shown.
On the right side there is shown a sum of ${\rm SU}(2|1)$ representations specified by their spin values $s$ and coinciding with those found in \cite{FIS17}
for the relevant spin. For the considered simplest case of $q=1/2$, spin $s$ takes only one value, $s=1$\,.}

\label{figure1}
\end{figure}
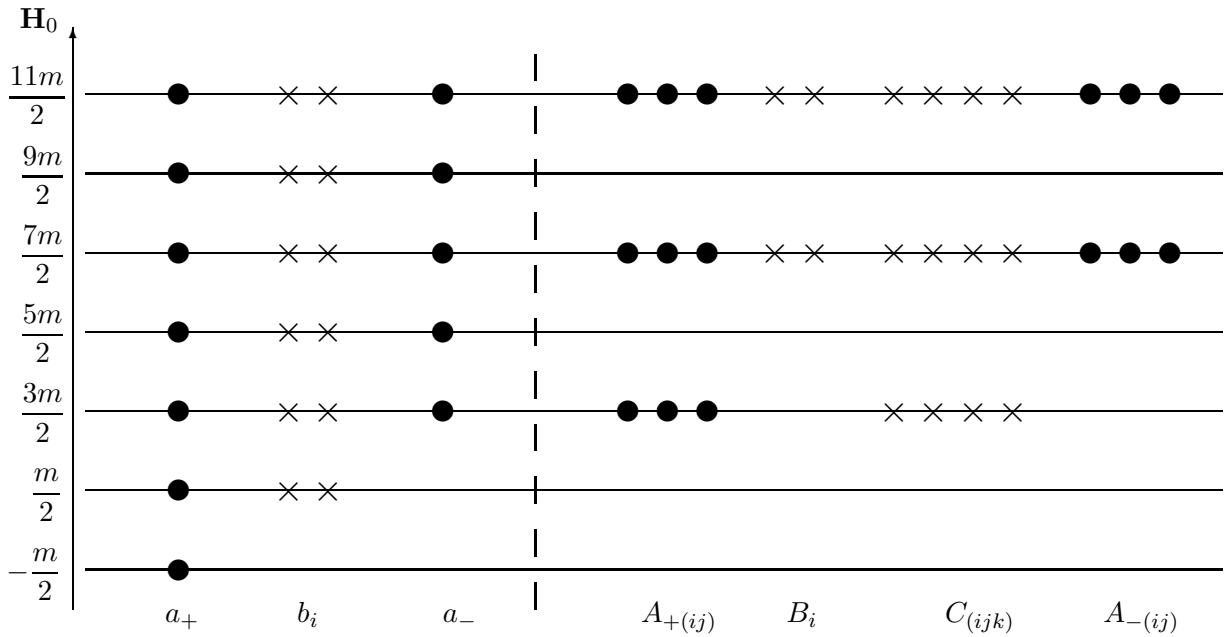

The $q=1$ case encompasses the same states as for $q=1/2$ ($s=1$) depicted in Fig. 1,
but also additional states with higher spins and higher energies.
The similar pictures  persist at lager $q$.

Summarizing the above discussion, we observe the basic distinction  between the one-particle system of ref. \cite{FIS17}
and the center-of-mass sector of $n$-particle system considered here. In the former case, the energy spectrum arises as a solution of the eigenvalue problems
of the type \p{eq-conf}, with the Hamiltonians involving a sum of the oscillator
and the inverse square potentials. In the latter case, the energy spectrum  contains as well pure oscillator
excitations due to the presence of the eigenvalue problems of the type \p{osc-wf-2}.

\vspace{0.5cm}

\centerline{\underline{$n=3$}}

\vspace{0.3cm}

In this case there are three constraints of the type (\ref{constr-wf-0n}) and for integer $q$ they lead to the following dependence
of the wave function on the spin variables
\bea
    \Phi_0^{(2q)} &=& \left[\left( z_a z_b\right)\left( z_a z_c\right)\left( z_b z_c\right)\right]^{q} \phi +
\sum_{a\neq b, a\neq c, b\neq c}\left( z_a z_b\right)^{q-1} \left[\left( z_a z_c\right)\left(z_b z_c\right)\right]^{q}
z_a^{i} z_b^{k}\,\Phi_{ab}{}_{(ik)}+ \ldots  \nn
    && +\,z_1^{i_{1}}\ldots z_1^{i_{2q}} z_2^{j_{1}}\ldots z_2^{j_{2q}} z_3^{k_{1}}\ldots z_3^{k_{2q}}
\Phi_{(i_{1}\ldots i_{2q} j_{1}\ldots j_{2q} k_{1}\ldots k_{2q})}\,.\label{expand-wf-01-3}
\eea
The component wave functions in the expansion (\ref{expand-wf-01-3}) are functions of $x_0$ and ${\psi}^i_{0}$, and they display
the dependence on ${\psi}^i_{0}$ similar to that in (\ref{expand-wf-011}).

In the three-spinor case, the relations analogous to (\ref{eq-1-2}), (\ref{eq-2-2}), (\ref{eq-3-2}), (\ref{eq-4-2}) are also valid,
the difference is that now an additional spin variable  $z_3^i$ appears in the products.
As a result, in the energy spectrum we find the same states  as in the $n=2$ case, though with a bigger multiplicity (due to extra indices $ab$ in (\ref{expand-wf-01-3})),
as well as the states of higher spins due to the presence of the additional spin variable  $z_3^i$.

In the case of half-integer $q$, the $n=3$ wave function has an expansion in which the component wave functions carry odd numbers of
spinor indices, as opposed to the expansion (\ref{expand-wf-01-3}). For example,
for $q=1/2$ wave function is
\be
\Phi_0^{(1)} = \left( z_2 z_3\right)z_1^{i}\,\Phi_{1\, i} +
\left( z_1 z_3\right)z_2^{i}\,\Phi_{2\, i} + \left( z_1 z_2\right)z_3^{i}\,\Phi_{3\, i} \,. \label{expand-wf-02-3}
\ee
The superwave functions  $\Phi_{a\, i}(x_0,\psi_0)$ display the energy spectrum of the one-particle system of ref. \cite{FIS17},
this time with the three-fold degeneracy.

\vspace{0.3cm}

The pictures for higher $n$ are similar to those for $n=2$ and $n=3$, such
that the number of states and the values of admissible spins are increasing at increasing $n$.

\section{Calogero--Moser system without center-of-mass sector}

As was mentioned in Introduction, there are two methods of finding the quantum energy spectrum of
multiparticle Calogero-type systems: either by considering matrix models which produce
physically equivalent Calogero-type systems after gauge-fixing and the corresponding reduction of phase space,
or through introducing Dunkl operators and passing to a generalized oscillator system.
In this section we apply the first method to quantize the $\mathcal{N}\,{=}\,4$
spin Calogero--Moser model under consideration in the matrix formulation, with the center-of-mass sector detached (Sect. 6.1).
The case of the reduced-phase space is briefly addressed in Sect. 6.2.

\subsection{Quantization in matrix formulation}
We consider the quantization of the matrix subsystem
in which $\widehat{su}(2|1)$ superalgebra is formed
by the generators ($\hat{\mathbf{Q}}^k$, $\hat{\bar{\mathbf{Q}}}_k$,
$\hat{\bf{H}}$, $\hat{\bf{I}}^i_k$, $\hat{\bf{F}}$) defined in (\ref{S-matr}), (\ref{def-Hn-matr}), (\ref{def-In-matr})
and (\ref{def-Fn-matr}).
The basic operators of this system are
spin operators ${\mathbf Z}_a^i$, $\bar {\mathbf Z}^a_i$ and traceless matrix operators
$\hat{\mathbf{X}}_a{}^b$, $ \hat{\mathbf{P}}_a{}^b$, $\hat{\bm{\Psi}}_a{}^b$, $\hat{\bar{\bm{\Psi}}}_a{}^b$
subject to the constraints (\ref{T-q-tr}) (as usual, applied to the physical states).

Introducing creation and annihilation even operators
\bea\label{AA+-def}
&&{\mathbf A}_a{}^b =  \frac{1}{\sqrt{2m}}\left(\hat{\mathbf P}_a{}^b -im\hat{\mathbf X}_a{}^b\right),
\quad {\mathbf A}^+_a{}^b =  \frac{1}{\sqrt{2m}}\left(\hat{\mathbf P}_a{}^b +im\hat{\mathbf X}_a{}^b\right),
\\
&&{\rm Tr}\left({\mathbf A}\right)={\rm Tr}\left({\mathbf A}^{+}\right)=0\,, \nonumber
\eea
we rewrite the Hamiltonian \eqref{def-Hn-matr} in the form
\be
\label{Ham-qn-matr-m-A}
\hat{\bf{H}}= \frac{m}{2}\, {\rm Tr}\left\{{\mathbf A}^+,{\mathbf A} \right\}
+ \frac{m}{2}\, {\rm Tr}\left[\hat{\bm{\Psi}}^k,\hat{\bar{\bm{\Psi}}}_k\right].
\ee
In this notation,  the supercharges (\ref{S-matr}) are rewritten as
\be\label{cr-an-Q}
\hat{\mathbf{Q}}^k = \sqrt{2m}\,{\rm Tr}\left({\mathbf A}\hat{\bm{\Psi}}^k\right),\qquad
\hat{\bar{\mathbf{Q}}}_k =  \sqrt{2m}\,{\rm Tr}\left({\mathbf A}^{+}\hat{\bar{\bm{\Psi}}}_k\right),
\ee
where quantum (anti)commutators of the involved operators are
\bea\label{qu-A-Psi}
\left[{\mathbf A}_a{}^b,{\mathbf A}^+_c{}^d \right]=\delta_a{}^d \delta_c{}^b-\frac{1}{n}\,\delta_a{}^b \delta_c{}^d,\qquad
\left\lbrace\hat{\bm{\Psi}}^i{}_a{}^b,\hat{\bar{\bm{\Psi}}}_j{}_c{}^d \right\rbrace=\left(\delta_a{}^d \delta_c{}^b-\frac{1}{n}\,\delta_a{}^b \delta_c{}^d\right)\delta^i_{j}\,.
\eea
The constraints (\ref{T-q-tr}) take the form
\be\label{T-q-tr-m-A}
{\mathbf G}_a{}^b =  \left[ {\mathbf A}^+,{\mathbf A} \right]_a{}^b +\left\{\hat{\bar{\bm{\Psi}}}_k,\hat{\bm{\Psi}}^k \right\}_a{}^b
+ {\mathbf Z}_a^k \bar {\mathbf Z}^b_k - \left(2q+n-\frac{1}{n}\right)\delta_a{}^b \simeq 0\,.
\ee
They involve the spin operators, with the non-vanishing commutator
\be\label{qu-ZbZ}
\left[{\mathbf Z}_a^i, \bar{\mathbf Z}^b_k \right]= -\,\delta_{k}^i \delta_{a}^b \,.
\ee

The operators ${\mathbf Z}_a^i$, ${\mathbf A}^+_a{}^b$, $\hat{\bm{\Psi}}^i{}_a{}^b$ form a full set of  creation operators.
Therefore, the general structure of the physical states is as follows
\be\label{phys-st-matr-gen}
{\mathbf Z}_{a_1}^{i_1}\ldots {\mathbf Z}_{a_{k_1}}^{i_{k_1}}
{\mathbf A}^+_{b_{1}}{}^{c_{1}}\ldots {\mathbf A}^+_{b_{k_2}}{}^{c_{k_2}}
\hat{\bm{\Psi}}^{j_{1}}{}_{d_{1}}{}^{e_{1}}\ldots \hat{\bm{\Psi}}^{j_{k_3}}{}_{d_{k_3}}{}^{e_{k_3}}\left.|0\right\rangle.
\ee
In the holomorphic realization,
\bea\label{hol-real-gen}
    &&{\mathbf Z}_{a}^{i}={z}_{a}^{i}\,,\qquad \bar{\mathbf Z}^{a}_{i}=\frac{\partial}{\partial {z}_{a}^{i}}\,,\nn
    &&{\mathbf A}^+_{b}{}^{c}=\hat{a}^+_{b}{}^{c}=a^+_{b}{}^{c} - \frac{1}{n}\,\delta_{b}^{c}\,a^+_{d}{}^{d}\,,\qquad
    {\mathbf A}_{b}{}^{c}=\frac{\partial}{\partial \hat{a}^+_{c}{}^{b}}=\frac{\partial}{\partial a^+_{c}{}^{b}}-\frac{1}{n}\,\delta_{b}^{c}\,\frac{\partial}{\partial a^+_{d}{}^{d}}\,,\nn
    &&\hat{\bm{\Psi}}^{i}{}_{b}{}^{c}=\hat{{\Psi}}^{i}{}_{b}{}^{c}=\Psi^i{}_{b}{}^{c} - \frac{1}{n}\,\delta_{b}^{c}\,
\Psi^i{}_{d}{}^{d}\,,\qquad
\hat{\bar{\bm{\Psi}}}_{i}{}_{b}{}^{c}=\frac{\partial}{\partial{\hat{\Psi}}^{i}{}_{c}{}^{b}}=
\frac{\partial}{\partial{\Psi}^{i}{}_{c}{}^{b}}- \frac1{n}\,\delta_{b}{}^{c}\frac{\partial}{\partial{\Psi}^{i}{}_{d}{}^{d}}\,,
\eea
we deal with the traceless objects $\hat{a}^+$ and $\hat{\Psi}^i$. Then the physical states \eqref{phys-st-matr-gen} are rewritten as
\be
z_{a_1}^{i_1}\ldots z_{a_{k_1}}^{i_{k_1}}
{\hat a}^+_{b_{1}}{}^{c_{1}}\ldots {\hat a}^+_{b_{k_2}}{}^{c_{k_2}}
\hat{\Psi}^{j_{1}}{}_{d_{1}}{}^{e_{1}}\ldots \hat{\Psi}^{j_{k_3}}{}_{d_{k_3}}{}^{e_{k_3}}\left.|0\right\rangle.
\ee
The constraint (\ref{T-q-tr-m-A}) indicates that all physical states are singlets of ${\mathrm{SU}(n)}$
(see \cite{Poly-gauge,Poly2001,HellRaa,Poly-rev}). This is also a direct consequence of vanishing of all Casimir operators
on the states \p{phys-st-matr-gen}:
\be\label{Casim}
{\mathbf G}_a{}^b {\mathbf G}_b{}^c\ldots {\mathbf G}_e{}^a \simeq  0\,.
\ee
On the other hand, the states \eqref{phys-st-matr-gen} belong to irreducible representations  of the group $\mathrm{SU}(2)$ with the generators $\mathbf{S}_{(ij)}$
defined in \p{S-quant} and the group $\mathrm{SU}(2)\times \mathrm{U}(1)$ with the generators \eqref{def-In-matr}, \eqref{def-Fn-matr} acting only on fermionic fields.

Eigenvalues of the Hamiltonian (\ref{Ham-qn-matr-m-A}) on the states (\ref{phys-st-matr-gen})
are specified by the numbers $N_{\mathbf A}$ and $N_{\mathbf\Psi}$ of the operators ${\mathbf A}^+_{b}{}^{c}$ and $\hat{\bm{\Psi}}^{i}{}_{b}{}^{c}$:
\be\label{wf-matr-gen-en}
E=m\left(N_{\mathbf A} + N_{\mathbf\Psi}-\frac{n^2-1}{2}\right).
\ee

Here we will basically limit our consideration to the pure bosonic case, without odd operators $\hat{\bm{\Psi}}^{i}{}_{b}{}^{c}$, $\hat{\bar{\bm{\Psi}}}_{j}{}_{b}{}^{c}$. The
set of fermionic states can be generated by action of the supercharges on the subset of bosonic states. Examples of fermionic states will be constructed
below for few simple particular cases.

The trace part of the constraints (\ref{T-q-tr-m-A}) leads to homogeneity of the physical states of degree $2qn$ with respect
to the spin operators ${\mathbf Z}\,$. In addition,
the property that physical states are the ${\mathrm{SU}(n)}$ singlets implies the following structure for them \cite{Poly-gauge,Poly2001,HellRaa,Poly-rev}
\bea\label{phys-st-matr-gen-bose-sing}
    \Phi^{(2q,s,\ell)}& \simeq &\left[{\rm Tr}\left({\mathbf A}^+\right)^2\right]^{p_2}\left[{\rm Tr}\left({\mathbf A}^+\right)^3\right]^{p_3}\ldots\left[{\rm Tr}\left({\mathbf A}^+\right)^n\right]^{p_n}\nn
    &&\times\prod^{2q-1}_{r=0}
\left\lbrace\varepsilon^{a_1a_2\ldots a_n}\left[\left({\mathbf A}^{+}\right)^{l_{nr+1}}{\mathbf Z}^{i_{nr+1}}\right]_{a_1}\ldots\left[\left({\mathbf A}^{+}\right)^{l_{nr+n}}{\mathbf Z}^{i_{nr+n}}\right]_{a_n}\right\rbrace\left.|0\right\rangle,
\eea
where $p_2,p_3\ldots, p_n$ are arbitrary integers and $0\leq l_{nr+1}\leq l_{nr+2}\ldots l_{nr+n}<n$. The wave function $\Phi^{(2q,s,\ell)}$ in \eqref{phys-st-matr-gen-bose-sing} is given up to the coefficients $C_{(i_1 i_2\ldots i_{2s})}$\,, where
the number $s\leq nq$ can be interpreted as ${\mathrm{SU}(2)}$ spin,
with $2s$ being the number of symmetrized ${\mathrm{SU}(2)}$ indices of spin variables.
It is worth pointing out that for $l_r<n/2$ the coincident degrees, $l_r=l_p$, are permitted.
Besides, such a degree cannot appear more than once in the products of monomials
\be
    \left\lbrace\varepsilon^{a_1a_2\ldots a_n}\left[\left({\mathbf A}^{+}\right)^{l_{nr+1}}{\mathbf Z}^{i_{nr+1}}\right]_{a_1}\ldots\left[\left({\mathbf A}^{+}\right)^{l_{nr+n}}{\mathbf Z}^{i_{nr+n}}\right]_{a_n}\right\rbrace,\qquad r=0,1\ldots 2q-1\,.
\ee
For the highest spin $s=nq$, the degrees are given by
\bea
    l_{nr+1} = 0\,,\quad l_{nr+2}=1\,,\quad\ldots\quad l_{nr+n}=n-1\,,\qquad r=0,1\ldots 2q-1\,.\label{s=nq}
\eea

Degeneracy analysis of bosonic wave functions for the quantum spin Calogero model was considered in \cite{FLP2013}, where it was noticed that some of possible spin states may vanish.
Here we consider the matrix construction for the system with the center-of-mass sector detached,
\footnote{The construction of \cite{FLP2013} implies reduction to the angular spin Calogero model by separating the radial coordinate,
which corresponds in the quantum case to setting $p_2=0$ in \eqref{phys-st-matr-gen-bose-sing}.} where some of these spin states may also vanish.
Listing all admissible degree numbers $l_{nr+1},l_{nr+2}\ldots, l_{nr+n}$ is a rather complicated task.

On the states (\ref{phys-st-matr-gen-bose-sing}) the energy (\ref{wf-matr-gen-en}) take the values
\be
E=m\left(\sum_{k=2}^{n}k p_k + \sum_{k=1}^{2qn} l_k -\frac{n^2-1}{2}\right)\,.\label{energy}
\ee
The energy is maximal for the choice \eqref{s=nq}:
\bea
    E_{(s=nq)}=m\left(\sum_{k=2}^{n}k p_k +\left(n-1\right)nq -\frac{n^2-1}{2}\right).
\eea
The minimal energy corresponds to the choice $p_2=p_3\ldots p_n=0$ and
$l_{nr+1}=l_{nr+2}=0$, $l_{nr+3}=l_{nr+4}=1$, $l_{nr+5}=l_{nr+6}=2$, etc:
\bea
    &&E_{\rm min}=m\left[n\left(\frac{n}{2}-1\right)q -\frac{n^2-1}{2}\right],\quad {\rm for\;even}\;n\,,\nn
    &&E_{\rm min}=m\left[\frac{\left(n-1\right)^2 q}{2} -\frac{n^2-1}{2}\right],\quad {\rm for\;odd}\;n>1\,.
\eea

The fermionic states are constructed with the help of the operators $\hat{\bm{\Psi}}^{i}{}_{b}{}^{c}$, on the pattern of (\ref{phys-st-matr-gen-bose-sing}).
Such physical states have additional contributions $m N_{\mathbf\Psi}$ to the energy value \eqref{energy}. As was already mentioned,
full wave functions can be generated from the bosonic states \eqref{phys-st-matr-gen} by acting on them by the supercharges \eqref{cr-an-Q}.
Casimir operators \eqref{su21-Cas2}, \eqref{su21-Cas2} take the following values on the states \eqref{phys-st-matr-gen}
 and those produced from   \eqref{phys-st-matr-gen} by ${\rm SU}(2|1)$ supersymmetry transformation:
\bea
    &&m^2{\mathbf C}_2 = \left(E+\frac{\left(n^2-1\right)m}{2}\right)\left(E+\frac{\left(n^2-3\right)m}{2}\right),\nn
    &&m^3{\mathbf C}_3 = \left(E+\frac{\left(n^2-2\right)m}{2}\right){\mathbf C}_2\,.
\eea
Casimirs can take zero eigenvalues only for $n=2$ at arbitrary $q$ and for $n=3$ at $q=1/2$ (we consider $q>0$ in this paper). The corresponding sets
of the quantum states belong to atypical representations of ${\rm SU}(2|1)$. For illustration, we will consider here these two cases in some detail.

\vspace{0.5cm}

\centerline{\underline{$n=2$}}

\vspace{0.3cm}
In this case the Hamiltonian is written as
\be\label{H-noc-2}
\hat{\bf{H}}= m\,{\rm Tr}\left({\mathbf A}^+{\mathbf A}\right)+m\, {\rm Tr}\left(\hat{\bm{\Psi}}^k\hat{\bar{\bm{\Psi}}}_k\right) - \frac{3m}{2}\,.
\ee
Bosonic wave functions, from which the full set of the wave functions can be produced by the supercharges (\ref{qu-A-Psi}), are given by
\bea
    \Phi^{(2q,s,\ell)}=\left[{\rm Tr}\left({\mathbf A}^+\right)^2\right]^{\ell}\left(\varepsilon_{ij}\varepsilon^{ab}{\mathbf Z}_a^i{\mathbf Z}_b^j\right)^{2q-s}
    {\mathbf Z}_{a_1}^{i_1}{\mathbf Z}_{a_2}^{i_2}\ldots{\mathbf Z}_{a_{2s}}^{i_{2s}}\prod_{k=1}^s\varepsilon^{a_kb_k}{\mathbf A}^{+}_{b_k}{}^{a_{s+k}}C_{(i_1 i_2\ldots i_{2s})}\left.|0\right\rangle,
\eea
where $C_{(i_1 i_2\ldots i_{2s})}$ are coefficients with $2s$ symmetric indices.
The wave functions $\Phi^{(2q,s,\ell)}$ are eigenfunctions of the Hamiltonian (\ref{H-noc-2}),
\be
    \hat{\bf{H}}\Phi^{(2q,s,\ell)} = E_{(s,\ell)}\Phi^{(2q,s,\ell)},\qquad s=0,1\ldots 2q\,,
\ee
with the energy eigenvalues
\be\label{matr_energy}
E_{(s,\ell)} = m\left(2\ell + s - \frac{3}{2}\right).
\ee
The  wave functions on which Casimirs take zero values, {\it i.e.}, those  belonging to atypical representations of $\mathrm{SU}(2|1)$,
correspond to the choice $s=0$\,, $\ell=0$:
\bea
    \Phi^{\left(2q,0,0\right)}=\left(\varepsilon_{ij}\varepsilon^{ab}{\mathbf Z}_a^i{\mathbf Z}_b^j\right)^{2q}\left.|0\right\rangle,
\eea
This ground state wave function is $\mathrm{SU}(2|1)$ singlet, since it is annihilated by both supercharges.
There is still another atypical non-singlet bosonic state corresponding to $s=1$\,, $\ell=0$:
\be
\Phi^{\left(2q,1,0\right)}=\left(\varepsilon_{k_1 k_2}\varepsilon^{c_1 c_2}{\mathbf Z}_{c_1}^{k_1}{\mathbf Z}_{c_2}^{k_2}\right)^{2q-1}
\varepsilon^{ab}{\mathbf Z}_{a}^{(i_1}{\mathbf Z}_{d}^{i_2)}{\mathbf A}^{+}_{b}{}^{d}\left.|0\right\rangle,\label{BosFund}
\ee
which gives rise to the fundamental $\mathrm{SU}(2|1)$ representation. The other two components of this representation
are generated from \eqref{BosFund} by  ${\rm SU}(2|1)$ supercharges:
\be
\hat{\mathbf{Q}}^j\Phi^{\left(2q,1,0\right)}=\sqrt{2m}\left(\varepsilon_{k_1 k_2}
\varepsilon^{c_1 c_2}{\mathbf Z}_{c_1}^{k_1}{\mathbf Z}_{c_2}^{k_2}\right)^{2q-1}\varepsilon^{ab}{\mathbf Z}_{a}^{(i_1}{\mathbf Z}_{d}^{i_2)}
\hat{\bm{\Psi}}^j{}_{b}{}^{d} \left.|0\right\rangle, \; {\rm etc}.\lb{BosFund1}
\ee

\vspace{0.5cm}

\centerline{\underline{$n=3, \;q=1/2$}}

\vspace{0.3cm}
The $n=3$ Hamiltonian reads
\be
\hat{\bf{H}}= \frac{m}{2}\, {\rm Tr}\left\{{\mathbf A}^+,{\mathbf A} \right\}
+ \frac{m}{2}\, {\rm Tr}\left[\hat{{\bm{\Psi}}}^k,{\hat{\bar{\bm{\Psi}}}}_k\right]=m\,{\rm Tr}\left({\mathbf A}^+{\mathbf A}\right)+m\,
{\rm Tr}\left(\hat{\bm{\Psi}}^k\hat{\bar{\bm{\Psi}}}_k\right) - 4m\,.
\ee
For $q=1/2$, the bosonic wave functions $\Phi^{(2q,s,\ell)}$ as eigenfunctions of this Hamiltonian are constructed as
\bea
    &&\Phi^{(1,1/2,\ell)}=\left[{\rm Tr}\left({\mathbf A}^+\right)^2\right]^{p_2}\left[{\rm Tr}\left({\mathbf A}^+\right)^3\right]^{p_3}\varepsilon^{a_1 a_2 a_3}\varepsilon_{ij}{\mathbf Z}^{i}_{a_1}{\mathbf Z}^{j}_{a_2}\left({\mathbf A}^{+}{\mathbf Z}^{k}\right)_{a_3} C_{k}\left.|0\right\rangle,\nn
    &&\Phi^{\prime(1,1/2,\ell)}=\left[{\rm Tr}\left({\mathbf A}^+\right)^2\right]^{p_2}\left[{\rm Tr}\left({\mathbf A}^+\right)^3\right]^{p_3}\varepsilon^{a_1 a_2 a_3}\varepsilon_{ij}\left({\mathbf A}^{+}{\mathbf Z}^{i}\right)_{a_1}\left({\mathbf A}^{+}{\mathbf Z}^{j}\right)_{a_2}{\mathbf Z}^{k}_{a_3} C^{\prime}_{k}\left.|0\right\rangle,\nn
    &&\Phi^{(1,3/2,\ell)}=\left[{\rm Tr}\left({\mathbf A}^+\right)^2\right]^{p_2}\left[{\rm Tr}\left({\mathbf A}^+\right)^3\right]^{p_3}\varepsilon^{a_1 a_2 a_3}{\mathbf Z}^{i}_{a_1}\left({\mathbf A}^{+}{\mathbf Z}^{j}\right)_{a_2}\left(\left[{\mathbf A}^{+}\right]^2{\mathbf Z}^{k}\right)_{a_3} C_{(ijk)}\left.|0\right\rangle,\nn
\eea
with the coefficients $C_{k}$\,, $C^{\prime}_{k}$\,, $C_{(ijk)}$\,.
They have the following energy values
\bea
    E_{\left(1/2,\ell\right)} = m\left(\ell -3\right),\qquad
    E^{\prime}_{\left(1/2,\ell\right)} = m\left(\ell -2\right),\qquad
    E_{\left(3/2,\ell\right)} = m\left(\ell -1\right),
\eea
where $\ell = 2p_2+3p_3$\,.
The minimal energy is achieved on the state
\bea
    \Phi^{(1,1/2,0)}=\varepsilon^{a_1 a_2 a_3}\varepsilon_{ij}{\mathbf Z}^{i_1}_{a_1}{\mathbf Z}^{i_2}_{a_2}\left({\mathbf A}^{+}{\mathbf Z}^{k}\right)_{a_3}\left.|0\right\rangle
    \lb{BosFund3}
\eea
 and it is equal to
\bea
    E_{\left(1/2,0\right)} = -3m\,.
\eea
Casimir operators take zero values on this state. The action of the ${\rm SU}(2|1)$ supercharge,
\bea
    \hat{\mathbf{Q}}^j\Phi^{(1,1/2,0)}=\varepsilon^{a_1 a_2 a_3}\varepsilon_{i_1 i_2}{\mathbf Z}^{i_1}_{a_1}{\mathbf Z}^{i_2}_{a_2}\left(\hat{\bm{\Psi}}^j
    {\mathbf Z}^{k}\right)_{a_3}\left.|0\right\rangle, \lb{FermFund3}
\eea
produces an additional fermionic state which, together with  \eqref{BosFund3} and one more bosonic state
generated by further action of supercharges on \eqref{FermFund3}, constitute an atypical fundamental $\mathrm{SU}(2|1)$ supermultiplet.

\subsection{Quantization of the reduced spinning Calogero--Moser system}

We briefly discuss quantization of the reduced spinning Calogero--Moser multi-particle system
without center-of-mass defined in Sect. 4.2.1\,.
More explicitly, we consider the two-particle case $n\,{=}\,2$\,.

Introducing the holomorphic realization
\bea
    &&{\mathbf Z}_{a}^{i}:={z}_{a}^{i}\,,\qquad \bar{\mathbf Z}^{a}_{i}:=\frac{\partial}{\partial {z}_{a}^{i}}\,,\qquad
    {\mathbf p}_{a} :=-i\partial_{a} = -i\,\frac{\partial}{\partial x_{a}}\,,\nn
    &&{\bm{\psi}}^{i}{}_{a}={\psi}^{i}{}_{a},\qquad
    \bar{\bm{\psi}}_{i\,a}=\frac{\partial}{\partial {\psi}^{i}{}_{a}},\qquad
    {\bm{\psi}}^{i}{}_{b}{}^{c}={\psi}^{i}{}_{b}{}^{c},\qquad
\bar{\bm{\psi}}_{i}{}_{b}{}^{c}=\frac{\partial}{\partial {\psi}^{i}{}_{c}{}^{b}}\,,
\eea
the differential realization of Hamiltonian \eqref{H-tot-ab} on physical states is given by
\bea
    \sum_{a<b}\left[\frac{1}{2n}\,
\left[-\left(\partial_a-\partial_b\right)^2 + m^2 \left(x_a -x_b\right)^2\right]
+\frac{g_{ab}}{\left(x_{a}-x_{b}\right)^2}\right]+{\rm const}\,.\label{CMH}
\eea
Here $g_{ab}$ are eigenvalues of the quantum operators $\frac{1}{2}\left\lbrace{\mathbf T}_a{}^b , {\mathbf T}_b{}^a\right\rbrace (a<b)$, and they correspond
to spin couplings of two interacting particles $x_a$ and $x_b$\,.
As was shown in Sect. \ref{4.2}, one can represent $g_{ab}$ on bosonic states via \eqref{S_a} as
\bea
    g_{ab} = -\,\mathbf{S}^{(ij)}_a\mathbf{S}_{b\,(ij)}+2q\left(q+1\right),\qquad a<b\,.
\eea
This model is restricted to positive integer values of $2q$ and is referred to as ``matrix model'' in \cite{Poly-rev}.
Arbitrary values of $2q$ can be achieved
by applying the exchange operator formalism  involving Dunkl operators (see, e.g., \cite{MP92}).
As was already mentioned, our ${\rm SU}(2|1)$ supersymmetric system
yields just the ${\rm U}(2)$ spin matrix model as its bosonic core.

In the simpler case $g_{ab}\,{=}\,\vartheta\left(\vartheta \mp 1\right)$\,, quantization was given in \cite{Poly-PRL,BHV,BHKV} via Dunkl operators defined as
\bea
    {\cal D}_a = \partial_a  + \sum_{a(\neq b)}\frac{\vartheta}{x_a-x_b}\left(1-K_{ab}\right),
\eea
where $K_{ab}$ is a permutation operator, $K_{ab}\,x_b\, {=}\, x_a\, K_{ab}$\,.
Below we consider the simplest case $n\,{=}\,2$\,, where $g_{12}=s\left(s+1\right)$ and the operator $K_{12}$ becomes Klein-type operator acting on the relative coordinate $x_1-x_2$ as
\bea
K_{12}\left(x_1 - x_2\right) = \left(x_2 - x_1\right)K_{12} = -\left(x_1 - x_2\right)K_{12}\,,\qquad \left(K_{12}\right)^2 = 1\,.\label{Klein}
\eea

Let us consider in details the two-particle system ($n=2$). It is described by the algebra of quantum operators
\be \label{x-q-al}
\begin{array}{c}
\left[{\mathbf x}, {\mathbf p} \right]=i\,,\qquad
\left[{\mathbf Z}_1^k, \bar{\mathbf Z}^1_j \right]= \left[{\mathbf Z}_2^k, \bar{\mathbf Z}^2_j \right]=-\,\delta_{j}^k \,, \\ [7pt]
\left\{{\bm{\psi}}^k, \bar{\bm{\psi}}_{j} \right\} = \delta_{j}^k\,,\qquad
\left\{{\bm{\psi}}^k{}_{1}{}^2, \bar{\bm{\psi}}_{j\,2}{}^1 \right\} = \left\{{\bm{\psi}}^k{}_{2}{}^1, \bar{\bm{\psi}}_{j\,1}{}^2 \right\} =\delta_{j}^k\,,
\end{array}
\ee
where
\be\label{x-12-q}
{\mathbf x}=\frac{1}{\sqrt{2}}\left({\mathbf x}_1-{\mathbf x}_2\right)\,,\qquad
{\mathbf p}=\frac{1}{\sqrt{2}}\left({\mathbf p}_1-{\mathbf p}_2\right)\,,
\ee
\be\label{ps-12-q}
{\bm{\psi}}^i =\frac{1}{\sqrt{2}}\left({\bm{\psi}}^i{}_1-{\bm{\psi}}^i{}_2\right) ,\qquad
\bar{\bm{\psi}}_{i}=\frac{1}{\sqrt{2}}\left(\bar{\bm{\psi}}_{i\,1}-\bar{\bm{\psi}}_{i\,2}\right).
\ee
Below we use the following realization for them
\be
\begin{array}{c}
{\displaystyle
{\mathbf x}=x\,,\quad {\mathbf p}=-i\frac{\partial}{\partial x}\,,\qquad
{\bm{\psi}}^i{}={\psi}^i \,,\quad  \bar{\bm{\psi}}_{i} = \frac{\partial}{\partial {\psi}^i}\,,}\\ [8pt]
{\displaystyle {\mathbf Z}^i_1 = z^i_1\,,\quad {\mathbf Z}^i_2 = z^i_2\,,\qquad
\bar{\mathbf Z}_j^1= \frac{\partial}{\partial z^j_1} \,, \quad\bar{\mathbf Z}_j^2= \frac{\partial}{\partial z^j_2} \,,}\\ [8pt]
{\displaystyle
{\bm{\psi}}^i{}_{1}{}^2={\psi}^i{}_{1}{}^2 \,, \quad  {\bm{\psi}}^i{}_{2}{}^1={\psi}^i{}_{2}{}^1 \,,\qquad
\bar{\bm{\psi}}_{i\,1}{}^2 = \frac{\partial}{\partial {\psi}^i{}_{2}{}^1}\,, \quad
\bar{\bm{\psi}}_{i\,2}{}^1 = \frac{\partial}{\partial {\psi}^i{}_{1}{}^2}\,.}
\end{array}
\ee
Here $x$, $z^i_a$ and ${\psi}^i$, ${\psi}^i{}_{a}{}^b$, $a=1,2$ are complex commuting and fermionic anticommuting  variables, respectively.

The Hamiltonian \eqref{H-tot-ab} without center of mass takes the form
\be \label{H-tot-ab-2}
{\mathbb{H}} =\frac{1}{2}
\left(-\frac{\partial^2}{\partial x^2} + m^2 x^2\right)
+ m\left(\psi^{k}\frac{\partial}{\partial \psi^{k}}
+ \psi^{k}{}_1{}^2 \frac{\partial}{\partial \psi^{k}{}_1{}^2}
+ \psi^{k}{}_2{}^1 \frac{\partial}{\partial \psi^{k}{}_2{}^1} -3\right)
+ \frac{\left\{{\mathbf T}_1{}^2,{\mathbf T}_2{}^1 \right\}}{ 4x^2}
\,.
\ee
where
\bea
{\mathbf T}_1{}^2 &=& z_1^k \frac{\partial}{\partial z_2^k} +
\sqrt{2}\, \psi^k \frac{\partial}{\partial \psi^k{}_2{}^1}
- \sqrt{2}\, \psi^k{}_1{}^2 \frac{\partial}{\partial \psi^k}\,, \nonumber\\
{\mathbf T}_2{}^1 &= &z_2^k \frac{\partial}{\partial z_1^k} -
\sqrt{2}\, \psi^k  \frac{\partial}{\partial \psi^k{}_1{}^2}
+ \sqrt{2}\,  \psi^k{}_2{}^1 \frac{\partial}{\partial \psi^k}\,. \label{T12-2}
\eea
The operators (\ref{T12-2}) act in the following way on the variables entering the wave function
\be \label{T12-act-2}
\begin{array}{rcl}
{\mathbf T}_1{}^2\,: &\quad& z_2^k \,\rightarrow\, z_1^k \,,\qquad {\psi}^k  \,\rightarrow\, -\sqrt{2}\, \psi^k{}_1{}^2 \,,
\qquad {\psi}^k{}_2{}^1 \,\rightarrow\, \sqrt{2}\,{\psi}^k\,;\\ [8pt]
{\mathbf T}_2{}^1\,: &\quad& z_1^k \,\rightarrow\, z_2^k \,,\qquad {\psi}^k  \,\rightarrow\, \sqrt{2}\, \psi^k{}_2{}^1 \,,
\qquad {\psi}^k{}_1{}^2 \,\rightarrow\, -\sqrt{2}\,{\psi}^k
\end{array}
\ee
and give zero, while acting on other variables. Thus, the operators ${\mathbf T}_1{}^2 {\mathbf T}_2{}^1$
and ${\mathbf T}_2{}^1 {\mathbf T}_1{}^2$ transform all components in the expansion of the wave function into
themselves, with some coefficients including the vanishing ones. Therefore, on all components
the Hamiltonian (\ref{H-tot-ab-2}) has the standard form with the oscillator and conformal potentials. As the result, we can find its energy spectrum.

This system is similar to the one we have considered in Sect. 5, but it has a wider set of fermionic fields.
The spin $s$ is associated with the diagonal external SU(2) group \eqref{S-quant}. In contrast to \eqref{I-tot-0}, the SU(2) subgroup \eqref{I-tot-ab} of SU(2$|$1)
acts only on fermionic fields, which gives a different degeneracy picture for the supergroup SU(2$|$1).

Let us consider pure bosonic wave functions. They are given by
\bea
    &&\Omega^{\left(2q\right)} = \sum_{\ell}^{\infty}\sum_{s=0}^{2q}\Omega^{\left(2q,s,\ell\right)},\nn
    &&\Omega^{\left(2q,s,\ell\right)}\left(x, z^i_1, z^j_2\right) = \left(z^k_1 z_{k2}\right)^{2q-s}z^{i_1}_1 z^{i_2}_1\ldots z^{i_s}_1 z^{i_{s+1}}_2\ldots z^{i_{2s}}_2 A^{\left(s,\ell\right)}_{(i_1 i_2 \ldots i_{2s})}
    \left(x\right),
\eea
and are subject to the constraints
\bea
    &&{\mathbf T}_1 \Omega^{\left(2q\right)} = {\mathbf T}_2 \Omega^{\left(2q\right)} = 2q\,\Omega^{\left(q\right)},\qquad 2q=1,2,3\ldots \,,\nn
    &&{\mathbf T}_1 = z_1^k \frac{\partial}{\partial z_1^k}
+ {\psi}^k{}_1{}^2 \frac{\partial}{\psi^k{}_1{}^2} - {\psi}^k{}_2{}^1 \frac{\partial}{\psi^k{}_2{}^1}\,,\nn
    &&{\mathbf T}_2 = z_2^k \frac{\partial}{\partial z_2^k} - {\psi}^k{}_1{}^2 \frac{\partial}{\psi^k{}_1{}^2} + {\psi}^k{}_2{}^1 \frac{\partial}{\psi^k{}_2{}^1} \,. \label{constr-wf-2}
\eea

The eigenvalue problem for the Hamiltonian \eqref{H-tot-ab-2} amounts to the equation
\bea
    \frac{1}{2}\left[-\frac{\partial^2}{\partial x^2} + m^2 x^2+\frac{s\left(s+1\right)}{x^2}-6m\right]A^{\left(s,\ell\right)}\left(x\right)_{(i_1 i_2 \ldots i_{2s})}
    = E_{(s,\ell)}\,A^{\left(s,\ell\right)}_{(i_1 i_2 \ldots i_{2s})},\label{H_A}
\eea
which is solved as \footnote{As was discussed in \cite{FIS17}, the equation \eqref{H_A} has an additional solution which was thrown away for $s>0$
due to the presence of singularities at $x=0$. Here this additional solution must be thrown away even for the case $s=0$, applying the same reasoning to the fermionic
expansion of the full wave function.}
\bea
    &&A^{\left(s,\ell\right)}\left(x\right)_{(i_1 i_2 \ldots i_{2s})}=C_{(i_1 i_2 \ldots i_{2s})}\,x^{s+1}\,L^{(s + 1/2)}_{\ell}\left(mx^2\right)\exp(-mx^2/2)\,,\qquad \ell=0,1,2,\ldots\,,\nn
    &&E_{\left(s,\ell\right)} = m\left(2\ell+ s - \frac{3}{2}\right).
\eea
The energy spectrum is consistent with the energy spectrum \eqref{matr_energy} calculated in the matrix formulation.

An alternative construction of wave functions can be given via creation and annihilation operators \cite{Poly-PRL,BHV,BHKV}. To solve the equation \eqref{H_A}, we take the ansatz:
\bea
    A^{\left(s,\ell\right)}_{(i_1 i_2 \ldots i_{2s})}=C_{(i_1 i_2 \ldots i_{2s})}\,x^{s+1}\,\Phi^{\left(s,\ell\right)}_{+},
\eea
where $\Phi^{\left(s,\ell\right)}_{+}$ is an even function of $x$\,. Introducing the Klein operator $K\,{:=}\,K_{12}$ \eqref{Klein} satisfying
\bea
    K^2=1\,,\qquad K x =-xK, \qquad K\partial_x = - \partial_x K, \qquad K \Phi^{\left(s,\ell\right)}_{+} = \Phi^{\left(s,\ell\right)}_{+},
\eea
we define the creation and annihilation operators $\mathbf{a}^{\pm}$ through the Dunkl operator ${\cal D}$:
\bea
    &&\mathbf{a}^{\pm} = \mp{\cal D}+mx\,,\qquad {\cal D}=\left[\partial_{x}+\frac{s+1}{x}\left(1-K\right)\right], \qquad K\mathbf{a}^{\pm}=-\,\mathbf{a}^{\pm}K\,.
\eea
Then eq. \eqref{H_A} is rewritten as
\bea
    &&\frac{1}{2}\left[-\frac{\partial^2}{\partial x^2} + m^2 x^2+\frac{s\left(s+1\right)}{x^2}-6m\right]A^{\left(s,\ell\right)}
    =x^{s+1}{\mathbb{H}}_{+}\Phi^{\left(s,\ell\right)}_{+},
\eea
where
\bea
    &&{\mathbb{H}}_{+} = \frac{1}{2}\,\mathbf{a}^{+}\mathbf{a}^{-}+\frac{m}{2}\left(2sK+2K-5\right),\qquad
    \left[{\mathbb{H}}_+ , \mathbf{a}^{\pm}\right]=\pm m \mathbf{a}^{\pm}.
\eea
Taking into account that $K \Phi^{\left(s,\ell\right)}_{+} = \Phi^{\left(s,\ell\right)}_{+}$, the function  $\Phi^{\left(s,\ell\right)}_{+}$ is expressed as
\bea
    \Phi^{\left(s,\ell\right)}_{+} = \left(\mathbf{a}^{+}\right)^{2\ell}\Phi^{\left(s,0\right)}_{+},\qquad \mathbf{a}^{-}\,\Phi^{\left(s,0\right)}_{+}=0\,,
\eea
and the associate energy spectrum is
\bea
    {\mathbb{H}}_{+}\Phi^{\left(s,\ell\right)}_{+}=E_{\left(s,\ell\right)}\Phi^{\left(s,\ell\right)}_{+},\qquad
    E_{\left(s,\ell\right)} = m\left(2\ell+ s - \frac{3}{2}\right),\qquad
    \ell=0,1,2,\ldots.\label{H_+E}
\eea

One can also choose an alternative ansatz
\bea
    A^{\left(s,\ell\right)}=x^{s}\,\Phi^{\left(s,\ell\right)}_{-},
\eea
satisfying
\bea
    K\Phi^{\left(s,\ell\right)}_{-}=-\,\Phi^{\left(s,\ell\right)}_{-}.
\eea
The construction of wave functions via the creation and annihilation operators will give the same solution for the energy spectrum as \eqref{H_+E}. We skip details of this construction
 which is similar to the previous one.

\setcounter{equation}{0}
\section{Quantization of the full system (center-of-mass plus \break relative-coordinate sectors)}

The energy spectrum of the unified system, which is a sum of the center-of-mass sector of Sect.\,5 and
the relative coordinate system of Sect.\,6, can be found as a tensorial product of the spectra of these two subsystems.

In the ungauged matrix formulation, the bosonic wave functions are a generalization of \eqref{phys-st-matr-gen-bose-sing} in the holomorphic realization \eqref{hol-real-gen}:
\bea\label{wf-matr-gen-bose-n}
    && \Phi^{(2q, s, l)} \sim f_{(j_1\ldots j_{2s})}(x_0) \left[ {\mathrm{Tr}}\left(\hat{a}^+\right)^2\right]^{p_2}\left[ {\mathrm{Tr}}\left(\hat{a}^+\right)^3\right]^{p_3}
\ldots \left[ {\mathrm{Tr}}\left(\hat{a}^+\right)^n\right]^{p_n} \nn
    &&\times\prod^{2q-1}_{r=0}
\left\lbrace\varepsilon^{a_1a_2\ldots a_n}\left[\left(\hat{a}^{+}\right)^{l_{nr+1}}z^{i_{nr+1}}\right]_{a_1}\ldots\left[\left(\hat{a}^{+}\right)^{l_{nr+n}}z^{i_{nr+n}}\right]_{a_n}\right\rbrace
\left.|0\right\rangle\,.
\eea
Their general structure is quite specified by the three requirements:

\begin{itemize}
\item The wave functions should be ${\rm U}(n)$ invariant as a consequence of the constraint \p{T-q} (or its equivalent form \p{T-q-tr}). This means that all ${\rm U}(n)$ indices $a$ should be contracted with the appropriate invariant tensors;
\item They should be of degree $2qn$ with respect to the whole set of spin variables in virtue of the constraint \p{T-q-0};
\item All free ${\rm SU}(2)$ indices of the spin variables should be symmetrized and contracted with the indices of $f_{(i_1 \ldots i_{2s})}(x_0)$. The energy spectrum
of admissible spins of these functions extends from $s=0$ to $nq$ (for $2nq$ even) and from $s=1/2$ to $nq$ (for  $2nq$ odd).
\end{itemize}

All the fermionic wave functions can be obtained by action of the total supercharges on \p{wf-matr-gen-bose-n}. The basic distinctions of the total system from the multi-particle system of Sect.\,6 concern
the realizations of the $\mathrm{SU}(2)$ symmetry appearing in the anti-commutators of the supercharges as an internal subgroup of $\mathrm{SU}(2|1)$.
In the system with the center-of-mass sector detached considered in Sect.\,6, this $\mathrm{SU}(2)$ symmetry is given by \eqref{I-tot-ab}, acts only on the fermionic operators and gives rise
just to degeneracy of the energy spectrum. In the total  system, the internal  $\mathrm{SU}(2)$ symmetry acts on the indices $i,j,\ldots$ of all components
of the wave functions (\ref{wf-matr-gen-bose-n}) and their fermionic completion.

Taking into account the analysis of the  previous section, we see that the problem of description of all states
in the unified case (the option \textbf{III)} in Sect. 4.3) for an arbitrary $n$ is rather complicated.
At the same time, we can directly determine, for all possible cases, the full energy spectrum simply by applying the methods of the previous sections.
Let us briefly describe the energy spectrum for the choice of $n=2$.

In this simplest case the matrix system is described by the Hamiltonian
\bea
{\textbf H} &=& \frac12 \left(-\frac{\partial^2}{\partial x_0{}^2} + m^2 x_0{}^2 -1\right)
+ m\,{\psi}^i_{0}\,\frac{\partial}{\partial {\psi}^i_{0}}
 +\frac{1}{x_0{}^2} \left( -\frac{1}{4}\,{\mathbf S}^{(ik)}{\mathbf S}_{ik}+
{\mathbf S}^{(ik)}{\psi}_{0\,i}\,\frac{\partial}{\partial {\psi}^k_{0}} \right)\nn
    &&+\,m\,{\rm Tr}\left({\mathbf A}^+{\mathbf A}\right)+m\, {\rm Tr}\left(\bm{\Psi}^k\bar{\bm{\Psi}}_k\right) - 2m\,.
\eea
The traceless part of the general constraints (\ref{T-q-tr-m-A}),
\be
{\mathbf G}_a{}^b = \frac{i}{n}\left[{\mathbf X}_0,{\mathbf P}_0 \right]\delta_a{}^b + \left[ {\mathbf A}^+,{\mathbf A} \right]_a{}^b +\left\{{\bar{\bm{\Psi}}}_k,{\bm{\Psi}}^k \right\}_a{}^b
+ {\mathbf Z}_a^k \bar {\mathbf Z}^b_k - \left(2q+n\right)\delta_a{}^b \simeq 0\,,
\ee
requires wave functions to be ${\rm SU}(n)$ scalars, while its trace part fixes the degree of homogeneity with respect to spin variables:
\be
\sum_a {\mathbf G}_a{}^a = 0\qquad\Rightarrow\qquad\sum_a{\mathbf Z}_a^k \bar {\mathbf Z}^a_k - 4q=0\,.
\ee.

The bosonic wave functions are constructed as
\bea
    \Phi^{(2q,0,\ell,\ell_0)}&=&\mbox{H}_{\ell_0}\left(x_0\right)e^{-\frac{mx_0{^2}}{2}}\left[{\rm Tr}\left({\mathbf A}^+\right)^2\right]^{\ell}
    \left(\varepsilon_{ij}\varepsilon^{ab}{\mathbf Z}_a^i{\mathbf Z}_b^j\right)^{2q}\left.|0\right\rangle,\\
    \Phi^{(2q,s,\ell,\ell_0)}&=&A^{(\ell_0)}_{(i_1 i_2\ldots i_{2s})}\left(x_0\right)\left[{\rm Tr}\left({\mathbf A}^+\right)^2\right]^{\ell}
    \left(\varepsilon_{ij}\varepsilon^{ab}{\mathbf Z}_a^i{\mathbf Z}_b^j\right)^{2q-s}{\mathbf Z}_{a_1}^{(i_1}{\mathbf Z}_{a_2}^{i_2}\ldots{\mathbf Z}_{a_{2s}}^{i_{2s})}\nn
    &&\times\prod_{k=1}^s\varepsilon^{a_kb_k}{\mathbf A}^{+}_{b_k}{}^{a_{s+k}} \left.|0\right\rangle,\qquad s=1,2\ldots 2q\,, \lb{AllBos}
\eea
and possess the energies
\bea
    &&E_{(0,\ell,\ell_0)} = m\left(2\ell + \ell_0 - 2\right),\nn
    &&E_{(s,\ell,\ell_0)} = 2m\left(\ell + \ell_0 + s -\frac{1}{2}\right),\qquad s=1,2\ldots 2q\,.
\eea
The complete set of the quantum states is recovered through the action of ${\rm SU}(2|1)$ supercharges on the complete set of these bosonic wave functions.

The generic case in the reduced phase space formulation will be considered elsewhere.

\setcounter{equation}0
\section{Concluding remarks and outlook}

In this paper, we presented the full quantum description  of the ${\rm SU}(2|1)$  supersymmetric multi-particle
Calogero--Moser system with spin variables. It was constructed by making use of the matrix formulation of this system.
Due to the presence of spin variables, the system under consideration involves internal spin degrees of freedom
and so provides $\mathcal{N}{=}4$ supersymmetrization of $\mathrm{U}(2)$ spin Calogero--Moser system, as opposed
to the systems considered in refs. \cite{BGL,GLP-2,GLP-3}.

We obtained the explicit expressions for the classical and quantum charges of the mass-deformed $\mathcal{N}{=}4$ supersymmetry
inherent to  the multiparticle system considered.
The crucial role in quantization of this system is played by the property that it became possible to single out the center-of-mass
subsector in the full system. This allowed us to separately explore the case of the center of mass and the case without  the center-of-mass variables.
Knowing the energy spectrum in these two cases immediately allows one to derive the energy spectrum of the total system.

We computed the energy spectrum, exploiting the matrix formulation of the $\mathcal{N}{=}4$ supersymmetric
$\mathrm{U}(2)$ spin Calogero--Moser system.
An alternative  way of quantizing such systems is to deal with the reduced system,
involving the dynamical position coordinates only. Such a method \cite{Poly-PRL,BHV,BHKV,Poly1999,Poly1998,Poly-rev}
(the ``operator method'' in the terminology by A.\,Polychronakos) widely uses the Dunkl operators for
building the oscillator-like phase space of the multi-particle Calogero-type systems. Some simple examples of
applying this equivalent method within the model considered here  were already discussed in Sect. 6.2.
In the next publication we are planning to develop, in full generality, the applications of the operator
method  to the systems with spin variables.
On this way we expect, in particular, to find out some new generalizations of the Dunkl operators
and obtain complete set of independent conserved quantities (integrals) for a rigorous proof of integrability.
One more direction for the future study is to construct and quantize multi-particle Calogero-type models with higher-rank
deformed supersymmetries of the kind ${\rm SU}(m|n)$ and to reveal their relationships with the integrable structures in  ${\cal N}{=}\,4$ super Yang-Mills theory,
e.g., along the lines of ref. \cite{Dabh,AP}.

One more interesting problem is to elucidate a possible hidden superconformal symmetry of the multi-particle system considered.
In the one-particle case, the corresponding quantum-mechanical (massive) system \cite{FI16} was found to possess such
a hidden $\mathcal{N}{=}4$ superconformal symmetry associated with the supergroup ${\rm OSp}(4|2)$ \cite{FIS17}. In the quantum domain,
the corresponding  superalgebra $osp(4|2)$  acts  as a spectrum-generating algebra.
The existence of an analogous extension of ${\rm SU}(2|1)$ symmetry in the multi-particle case is an open question. In general, one could expect
as well a hidden $D(2,1;\alpha)$ supersymmetry for which ${\rm OSp}(4|2)$ is a particular case corresponding to the choice $\alpha =-1/2$. However, this possibility would require,
from the very beginning, some nonlinear sigma model action for the superfields $\mathscr{X}_b^a$ in \p{4N-gau} and, respectively, for the bosonic fields ${X}_b^a$ in \p{4N-gau-bose-1}.
The choice of ${\rm OSp}(4|2)$ is the unique one consistent with free kinetic terms for the bosonic fields, as long as one insists on the
supercharges~\eqref{Q-charges-q} being linear in fermionic variables~\cite{KLech}. Allowing for supercharge terms cubic in the fermionic operators
will constrain their coefficient functions by the so-called WDVV equations~\cite{BGL,GLP-2,GLP-3,KLP-9,LST,KLech}.
It will be interesting to develop a superspace variant of this more general situation.

\section*{Acknowledgements}

\noindent
This research was partially supported by  the joint DFG project LE 838/12-2 and the Heisenberg-Landau program.
The work of S.F., E.I. and S.S.
was partially supported by the RFBR Grant No. 16-52-12012, Russian Science Foundation Grant No. 16-12-10306
and Russian Ministry of Education and Science grant, project No. 3.1386.201.
This article is based
upon work from COST Action MP1405 QSPACE, supported by COST (European
Cooperation in Science and Technology).

\end{document}